\pdfoutput=1
\documentclass[12pt]{article}
\usepackage{graphicx}
\usepackage{subcaption}
\usepackage{amssymb}
\usepackage{amsmath}
\usepackage{amsfonts}
\usepackage{multirow}

\usepackage{xcolor}
\usepackage{url}
\usepackage{ulem}
\usepackage{float}
\usepackage{float}

\usepackage{pifont}% http://ctan.org/pkg/pifont

\usepackage{soul}
\usepackage[english]{babel}
\usepackage[T1]{fontenc}
\usepackage{lmodern}
\usepackage[utf8]{inputenc}  % f??r Unix-Systeme
\usepackage{bbm}

\usepackage{hyperref}
\hypersetup{
  colorlinks   = true, %Colours links instead of ugly boxes
  urlcolor     = blue, %Colour for external hyperlinks
  linkcolor    = black, %Colour of internal links
  citecolor   = black %Colour of citations
}

\usepackage[numbers,sort&compress]{natbib}
\graphicspath{{plots/}}

\newcommand{\lsim}
{\;\raisebox{-.3em}{$\stackrel{\displaystyle <}{\sim}$}\;}

\newcommand\tb{\tan\beta}

\newcommand\LP{\left(}
\newcommand\RP{\right)}

\newcommand\ReDiag{\mathop{%
  \raise .5pt\hbox{[}%
  \widetilde{\mathrm{Re}}%
  \raise .5pt\hbox{]}}}
\newcommand\ReOffDiag{\mathop{%
  \raise .5pt\hbox{$\llbracket$}%
  \widetilde{\mathrm{Re}}%
  \raise .5pt\hbox{$\rrbracket$}}}

\newcommand\MW{M_W}
\newcommand\MZ{M_Z}
\newcommand\Mh{M_h}

\newcommand\MA{M_A}

\newcommand\Sn{\tilde\nu}
\newcommand\Sl{\tilde l}

\newcommand\Slpm{\tilde l^\pm}

\newcommand\Sel[1]{\tilde e_{#1}}
\newcommand\Smu[1]{\tilde \mu_{#1}}

\newcommand\mse[1]{m_{\Sel{#1}}}
\newcommand\msl[1]{m_{\Sl_{#1}}}

\newcommand\stau{\tilde \tau}

\newcommand\mL{m_{\tilde l_L}}
\newcommand\mR{m_{\tilde l_R}}

\newcommand\ino[1]{\tilde\chi_{#1}}

\newcommand\chapm[1]{\ino{#1}^\pm}

\newcommand\chap[1]{\ino{#1}^+}
\newcommand\cham[1]{\ino{#1}^-}
\newcommand\cha{\chapm}
\newcommand\mcha[1]{m_{\chapm{#1}}}

\newcommand\neu[1]{\ino{#1}^0}
\newcommand\mneu[1]{m_{\neu{#1}}}

\newcommand\refeq[1]{Eq.~(\ref{#1})}
\newcommand\refeqs[1]{Eqs.~(\ref{#1})}

\newcommand\refse[1]{Sect.~\ref{#1}}

\newcommand\citere[1]{Ref.~\cite{#1}}
\newcommand\citeres[1]{Refs.~\cite{#1}}

\newcommand{\CP}{{\cal CP}}
\newcommand{\cp}{{\CP}}

\newcommand{\tev}{\,\, \mathrm{TeV}}
\newcommand{\gev}{\,\, \mathrm{GeV}}
\newcommand{\mev}{\,\, \mathrm{MeV}}

\newcommand\MO{\texttt{MicrOMEGAs}}
\newcommand\CM{\texttt{CheckMATE}}

\newcommand\pb{\ensuremath{\,\mbox{pb}}}
\newcommand\fb{\ensuremath{\,\mbox{fb}}}
\newcommand\ab{\ensuremath{\,\mbox{ab}}}

\newcommand\iab{\ensuremath{\,\ab^{-1}}}

\newcommand\msmu[1]{m_{\tilde{\mu}_{#1}}}
\newcommand\mstau[1]{m_{\tilde{\tau}_{#1}}}

\newcommand{\br}{\text{BR}}

\newcommand{\sig}{\sigma}

\def\order#1{\ensuremath{{\cal O}(#1)}}
\def\reffi#1{\mbox{Fig.~\ref{#1}}}

\def\ga{\gamma}
\def\De{\Delta}

\def\gmin2{\ensuremath{(g-2)_\mu}}
\def\amu{\ensuremath{a_\mu}}
\def \met  {\mbox{${E\!\!\!\!/_T}$}}
\newcommand{\ssi}{\ensuremath{\sig_p^{\rm SI}}}

\definecolor{Orange}{named}{orange}
\definecolor{Purple}{named}{purple}
\definecolor{Lightblue}{cmyk}{0.9,0.1,0.1,0.3}
\definecolor{dgelborange}{cmyk}{0.,0.3,0.5, 0.}
\definecolor{Lila}{rgb}{0.5,0.,1}
\definecolor{Darkgreen}{rgb}{0.,.7,0.2}

\evensidemargin \oddsidemargin
\allowdisplaybreaks
\begin{document}
\thispagestyle{empty}

\def\thefootnote{\fnsymbol{footnote}}

\begin{flushright}
\mbox{}
IFT--UAM/CSIC--21-138, 
IPMU21-0083\\
%arXiv:2111.nnnnn [hep-ph]
\end{flushright}

%\vspace{0.3cm}

\begin{center}

{\large\sc 
{\bf \boldmath{\gmin2} and SUSY Dark Matter:\\[.5em]
Direct Detection and Collider Search Complementarity}}

\vspace{0.3cm}

{\sc
Manimala Chakraborti$^{1}$%
\footnote{email: mani.chakraborti@gmail.com}% 
, Sven Heinemeyer$^{2}$%
\footnote{email: Sven.Heinemeyer@cern.ch}%
, Ipsita Saha$^{3}$%
\footnote{email: ipsita.saha@ipmu.jp}\\[.5em]
and Christian Schappacher$^{4}$%
\footnote{email: schappacher@kabelbw.de}
}

\vspace*{.5cm}

{\sl
$^1${Astrocent, Nicolaus Copernicus Astronomical Center
of the Polish Academy of Sciences,
ul. Rektorska 4, 00-614 Warsaw, Poland\\}

\vspace*{0.1cm}

$^2$Instituto de F\'isica Te\'orica (UAM/CSIC), 
Universidad Aut\'onoma de Madrid, \\ 
Cantoblanco, 28049, Madrid, Spain

\vspace*{0.1cm}

$^3$Kavli IPMU (WPI), UTIAS, University of Tokyo, Kashiwa,
Chiba 277-8583, Japan
} 

\vspace*{0.1cm}

$^4$Institut f\"ur Theoretische Physik,
Karlsruhe Institute of Technology, \\
76128, Karlsruhe, Germany (former address)

\end{center}

\vspace*{0.1cm}

\begin{abstract}
\noindent
The electroweak (EW) sector of the Minimal Supersymmetric Standard Model
(MSSM) can account for variety of experimental data.
The EW particles with masses of a few hundred GeV evade the LHC
searches owing to their small production cross sections. 
Such a light EW sector can in particular explain the reinforced
$4.2\,\sig$ discrepancy between the experimental result for the
anomalous magnetic moment of the muon, \gmin2, and its Standard Model
(SM) prediction.  
The lightest supersymmetric particle (LSP), assumed to be the lightest
neutralino, $\neu1$, as a Dark Matter (DM) candidate
is furthermore in agreement with the observed limits on
the DM content of the universe. Here the
Next-to LSP (NLSP) serves as a coannihilation partner and is naturally
close in mass to the LSP. Such scenarios are also to a large extent in
agreement with negative results from Direct Detection (DD) experiments.
The DM relic density can fully be explained by a nearly pure bino or a
mixed bino/wino LSP. Relatively light wino and higgsino DM, on the
other hand, remains easily below the DM relic density upper bound.
Using the improved limits on \gmin2, we explore the mass ranges of the
LSP and the NLSP in their correlation with the DM relic density for
bino, bino/wino, wino and higgsino DM. In particular analyze the
sensitivity of future DM DD experiments to these DM scenarios.
We find that higgsino, wino and one type of bino scenario can be
covered by future DD experiments. Mixed bino/wino and another type of
bino DM can reach DD cross sections below the neutrino floor. In these 
cases we analyze the complementarity with the (HL-)LHC and
future $e^+e^-$ linear colliders. We find that while the prospects
for the HL-LHC are interesting, but not conclusive, an $e^+e^-$ collider
with $\sqrt{s} \lsim 1 \tev$ can cover effectively all points of the
MSSM that may be missed by DD experiments.
\end{abstract}

%\pacs{}

\def\thefootnote{\arabic{footnote}}
\setcounter{page}{0}
\setcounter{footnote}{0}

\newpage

%%%%%%%%%%%%%%%%%%%%%%%%%%%%%%%%%%%%%%%%%%%%%%%%%%%%%%%%%%%%%%%%%%%%%%%%%%%%%%%
%%%%%%%%%%%%%%%%%%%%%%%%%%%%%%%%%%%%%%%%%%%%%%%%%%%%%%%%%%%%%%%%%%%%%%%%%%%%%%%

\section{Introduction}
\label{sec:intro}

Searches for Dark Matter (DM) is one of the main objectives in today's
particle and astroparticle physics. Searches at the LHC (or other
collider experiments) are complementary to the searches in ``direct
detection'' (DD) experiments. 
Among the Beyond the Standard Model (BSM) theories that predict a
viable DM particle the Minimal Supersymmetric Standard Model  
(MSSM)~\cite{Ni1984,Ba1988,HaK85,GuH86} is one of the leading candidates.
Supersymmetry (SUSY) predicts two scalar partners for all Standard
Model (SM) fermions as well as fermionic partners to all SM bosons. 
Furthermore, contrary to the SM case, the MSSM requires two Higgs doublets.
This results in five physical Higgs bosons instead of the single Higgs
boson in the SM: the light and heavy $\cp$-even Higgs bosons, 
$h$ and $H$, the $\cp$-odd Higgs boson, $A$, and the charged Higgs bosons,
$H^\pm$.
The neutral SUSY partners of the (neutral) Higgs and electroweak (EW) gauge
bosons gives rise to the four neutralinos, $\neu{1,2,3,4}$.  The corresponding
charged SUSY partners are the charginos, $\cha{1,2}$.
The SUSY partners of the SM leptons and quarks are the scalar leptons
and quarks (sleptons, squarks), respectively.
The lightest SUSY particle (LSP) is naturally the lightest neutralino,
$\neu1$. It can make up the full DM content of the
universe~\cite{Go1983,ElHaNaOlSr1984}, or, depending on its nature
only a fraction of it. In the latter case, an additional DM component could
be, e.g., a SUSY axion~\cite{Bae:2013bva},
which would then bring the total DM density into agreement with the
experimental measurement.

In \citeres{CHS1,CHS2,CHS3} we performed a comprehensive analysis of the EW
sector of the MSSM, taking into account all
relevant theoretical and experimental constraints. 
The experimental results comprised the direct searches at the
LHC~\cite{ATLAS-SUSY,CMS-SUSY}, the DM relic abundance~\cite{Planck}
(either as an upper limit~\cite{CHS2} or as a direct
measurement~\cite{CHS1,CHS3}), 
the DM direct detection (DD) experiments~\cite{XENON,LUX,PANDAX} and in
particular the  deviation of the anomalous magnetic moment
of the muon (either the previous result~\cite{CHS1,CHS2}, or the new,
stronger limits~\cite{CHS3}). 
Five different scenarios were analyzed, classified by the mechanism
that brings the LSP relic density into agreement with the measured
values. The scenarios differ by the Next-to-LSP (NLSP), or
equivalently by the mass hierarchies between the mass scales
determining the neutralino, chargino and slepton masses.
These mass scales are the gaugino soft-SUSY breaking parameters $M_1$
and $M_2$, the Higgs mixing parameter $\mu$ and the slepton soft
SUSY-breaking parameters $\msl{L}$ and $\msl{R}$, see \refse{sec:model}
for a detailed description. 
The five scenarios can be summarized as follows~\cite{CHS1,CHS2,CHS3}:
\begin{itemize}
\item[(i)]
higgsino DM ($\mu < M_1, M_2, \msl{L}, \msl{R}$),
DM relic density is only an upper bound (the rull relic density implies
$\mneu1 \sim 1 \tev$ and \gmin2\ cannot be fulfilled), 
$m_{\rm (N)LSP} \lsim 500 \gev$ with $m_{\rm NLSP} - m_{\rm LSP} \sim 5 \gev$;
\item[(ii)]
wino DM ($M_2 < M_1, \mu, \msl{L}, \msl{R}$),
DM relic density is only an upper bound, (the rull relic density implies
$\mneu1 \sim 3 \tev$ and \gmin2\ cannot be fulfilled),
$m_{\rm (N)LSP} \lsim 600 \gev$ with $m_{\rm NLSP} - m_{\rm LSP} \sim 0.3 \gev$;
\item[(iii)]
bino/wino DM with $\cha1$-coannihilation ($M_1 \lsim M_2$),
DM relic density can be fulfilled, $m_{\rm (N)LSP} \lsim 650\, (700) \gev$;
\item[(iv)]
bino DM with $\Slpm$-coannihilation case-L ($M_1 \lsim \msl{L}$),
DM relic density can be fulfilled, $m_{\rm (N)LSP} \lsim 650\, (700) \gev$;
\item[(v)]
bino DM with $\Slpm$-coannihilation case-R ($M_1 \lsim \msl{R}$),
DM relic density can be fulfilled, $m_{\rm (N)LSP} \lsim 650\, (700) \gev$.
\end{itemize}

Recently the ``MUON G-2'' collaboration published the results of their
Run~1 data~\cite{Abi:2021gix}, which is within $0.8\,\sig$ in
agreement with  the older BNL result on \gmin2~\cite{Bennett:2006fi}.
The combined measurement yields a deviation from the SM prediction of
$\De\amu = (25.1 \pm 5.9) \times 10^{-10}$, corresponding to $4.2\,\sig$.
Imposing this limit on the MSSM parameter space allows to set {\it upper} 
limits on the EW sector.
Here it is interesting to note that the old lower $2\,\sig$ limit on
$\De\amu$, $\De\amu^{-2\,\sig{\rm , old}} = 12.9 \times 10^{-10}$, 
coincidentally agrees quite well with the new lower limit,
$\De\amu^{-2\,\sig} = 13.3 \times 10^{-10}$.
Consequently, the new combined $\amu$ result confirmed the {\it upper}
mass limits obtained with the old $\amu$ result at a higher confidence level. 
While in \citeres{CHS1,CHS2} the old deviation (i.e.\ without the new
``MUON G-2'' result) was used for scenarios (i)-(v), scenarios
(iii)-(v) have been updated with the new result in \citere{CHS3}.
Other evaluations within the framework of SUSY using the new combined
deviation $\De\amu$ can be found in \citeres{Endo:2021zal,Iwamoto:2021aaf,Gu:2021mjd,VanBeekveld:2021tgn,Yin:2021mls,Wang:2021bcx,Abdughani:2021pdc,Cao:2021tuh,Ibe:2021cvf,Cox:2021gqq,Han:2021ify,Heinemeyer:2021zpc,Baum:2021qzx,Zhang:2021gun,Ahmed:2021htr,Athron:2021iuf,Aboubrahim:2021rwz,Chakraborti:2021bmv,Baer:2021aax,Altmannshofer:2021hfu,Chakraborti:2021squ,Zheng:2021wnu,Jeong:2021qey,Li:2021pnt,Dev:2021zty,Kim:2021suj,Ellis:2021zmg,Zhao:2021eaa,Frank:2021nkq,Shafi:2021jcg,Li:2021koa,Aranda:2021eyn,Aboubrahim:2021ily,Nakai:2021mha,Li:2021cte,Li:2021xmw,Lamborn:2021snt,Fischer:2021sqw,Forster:2021vyz,Ke:2021kgy,Ellis:2021vpp,Athron:2021dzk,Chakraborti:2021ynm}.

\smallskip
In this letter we address the implications of the new result for
$\De\amu$ for the DM predictions in the five scenarios. In a first
step we will analyze the predictions for the DM relic density as a
function of the (N)LSP masses. Here, in scenarios (iii)-(v) we will
show the results both for DM fulfilling the relic density, as well as
taking the DM density only as an upper bound.
In scenarios (i) and (ii), we analyze the case where a fraction of DM
relic density is contributed by $\neu1$ while being in agreement with
the $\De\amu$ requirement.
In a second step we evaluate the prospects for future DD experiments
in these five scenarios. We show that higgsino, wino and bino case-R DM
can be covered by the future DD experiments. Mixed bino/wino DM and bino
case-L DM, on the other hand, can reach DD cross sections below the
neutrino floor for a significant amount of model parameter space, if
the DM relic density remains substantially below 
the Planck measurement. In this case direct searches at the HL-LHC and
possibly at a future linear $e^+e^-$ collider will be necessary to
fully cover these scenarios. While \citeres{Endo:2021zal,Iwamoto:2021aaf,Gu:2021mjd,VanBeekveld:2021tgn,Yin:2021mls,Wang:2021bcx,Abdughani:2021pdc,Cao:2021tuh,Ibe:2021cvf,Cox:2021gqq,Han:2021ify,Heinemeyer:2021zpc,Baum:2021qzx,Zhang:2021gun,Ahmed:2021htr,Athron:2021iuf,Aboubrahim:2021rwz,Chakraborti:2021bmv,Baer:2021aax,Altmannshofer:2021hfu,Chakraborti:2021squ,Zheng:2021wnu,Jeong:2021qey,Li:2021pnt,Dev:2021zty,Kim:2021suj,Ellis:2021zmg,Zhao:2021eaa,Frank:2021nkq,Shafi:2021jcg,Li:2021koa,Aranda:2021eyn,Aboubrahim:2021ily,Nakai:2021mha,Li:2021cte,Li:2021xmw,Lamborn:2021snt,Fischer:2021sqw,Forster:2021vyz,Ke:2021kgy,Ellis:2021vpp,Athron:2021dzk,Chakraborti:2021ynm}
study the $\De\amu$ implications in SUSY models, to our knowledge a
DM analysis, particularly in view of the future detection prospect, as
performed here, has not been done.  

%%%%%%%%%%%%%%%%%%%%%%%%%%%%%%%%%%%%%%%%%%%%%%%%%%%%%%%%%%%%%%%%%%%%
%%%%%%%%%%%%%%%%%%%%%%%%%%%%%%%%%%%%%%%%%%%%%%%%%%%%%%%%%%%%%%%%%%%%

\section {The electroweak sector of the MSSM}
\label{sec:model}

In our notation for the MSSM we follow exactly \citeres{CHS1}. Here we
restrict ourselves to a very short introduction of the relevant parameters and
symbols of the EW sector of the MSSM, consisting of charginos,
neutralinos and scalar leptons. For the scalar quark sector, we assume it to
be heavy and not to play a relevant role in our analysis. Throughout this
paper we also assume absence of $\CP$-violation, i.e.\ 
that all parameters are real.

The masses and mixings of the neutralinos are set (on top of SM
parameters) by the $U(1)_Y$ and $SU(2)_L$ 
gaugino masses, $M_1$ and $M_2$, the Higgs mixing parameter $\mu$, 
as well as the ratio of the two vacuum expectation values (vevs)
of the two Higgs doublets, $\tb := v_2/v_1$. 
After the diagonalization of the mass matrix the four eigenvalues
give the four neutralino masses $\mneu1 < \mneu2 < \mneu3 <\mneu4$.
Similarly, the masses and mixings of the charginos are set (on top of SM
parameters) by $M_2$, $\mu$ and $\tb$. 
The diagonalization of the mass matrix yields the two chargino-mass
eigenvalues $\mcha1 < \mcha2$.

For the sleptons, as in \citere{CHS1}, we have chosen common soft
SUSY-breaking parameters for all three generations.
The charged slepton mass matrix are given (on top of SM parameters) by
the diagonal soft SUSY-breaking parameters $\mL^2$ and $\mR^2$ and the
trilinear Higgs-slepton coupling $A_l$ ($l = e, \mu, \tau$), where the
latter are set to zero. Mixing between the ``left-handed'' and
``right-handed'' sleptons is only relevant for staus, where the
off-diagonal entry in the mass matrix is dominated by $-m_\tau \mu \tb$.
Consequently, for the first two generations, the mass eigenvalues can
be approximated as $\msl1 \simeq \mL, \msl2 \simeq \mR$ (assuming
small $D$-terms).
In general we follow the convention that $\Sl_1$ ($\Sl_2$) has the
large ``left-handed'' (``right-handed'') component, i.e.\ they are not
mass ordered.
Besides the symbols are equal for all three generations, $\msl1$ and
$\msl2$, we also use symbols for the scalar electron, muon and tau masses,
$\mse{1,2}$, $\msmu{1,2}$ and $\mstau{1,2}$.
The sneutrino and slepton masses are connected by the usual SU(2) relation.

Overall, the EW sector at the tree level
can be described with the help of six parameters: $M_1$, $M_2$, $\mu$,
$\tb$, $\mL$ and $\mR$. Throughout our analysis we 
assume $\mu, M_1, M_2 > 0 $. In \citere{CHS1} it was shown that
choosing these parameters positive covers the relevant parameter space
once the \gmin2\ results are taken into account
(see, however, the discussion in \citere{Baum:2021qzx}).

%%%%%%%%%%%%%%%%%%%%%%%%%%%%%%%%%%%%%%%%%%%%%%%%%%%%%%%%%%%%%%%%%%%%%%%%%%

\medskip
Following the experimental limits from the
LHC~\cite{ATLAS-SUSY,CMS-SUSY} for strongly interacting particles, 
we assume that the colored sector of the MSSM is substantially heavier
than the EW sector, and thus does not play a role in our analysis. For the
Higgs-boson sector we assume that the radiative corrections to the light
$\cp$-even Higgs boson, which largely originate from the top/stop
sector, yield a value in agreement with the experimental data,
$\Mh \sim 125 \gev$. This yields stop masses naturally in the TeV
range~\cite{Bagnaschi:2017tru,Slavich:2020zjv}, in agreement 
with the LHC bounds. Concerning the heavy Higgs-boson mass scale, as
given by $\MA$, the $\cp$-odd Higgs-boson mass, we have shown
in \citeres{CHS1,CHS2,CHS3} that $A$-pole annihilation is largely
excluded. Consequently, we simply assume $\MA$ to be sufficiently
large to not play a role in our analysis.

%%%%%%%%%%%%%%%%%%%%%%%%%%%%%%%%%%%%%%%%%%%%%%%%%%%%%%%%%%%%%%%%%%%%%%%%%%
%%%%%%%%%%%%%%%%%%%%%%%%%%%%%%%%%%%%%%%%%%%%%%%%%%%%%%%%%%%%%%%%%%%%%%%%%%

\section {Relevant constraints}
\label{sec:constraints}

The SM prediction of \amu\ is given by~\cite{Aoyama:2020ynm}
(based on \citeres{Aoyama:2012wk,Aoyama:2019ryr,Czarnecki:2002nt,Gnendiger:2013pva,Davier:2017zfy,Keshavarzi:2018mgv,Colangelo:2018mtw,Hoferichter:2019mqg,Davier:2019can,Keshavarzi:2019abf,Kurz:2014wya,Melnikov:2003xd,Masjuan:2017tvw,Colangelo:2017fiz,Hoferichter:2018kwz,Gerardin:2019vio,Bijnens:2019ghy,Colangelo:2019uex,Blum:2019ugy,Colangelo:2014qya}
)%
\footnote{
In \citere{CHS1} a slightly different value was used, with a
negligible effect on the results.
}%
, 
\begin{align}
\amu^{\rm SM} &= (11 659 181.0 \pm 4.3) \times 10^{-10}~.
\label{gmt-sm}
\end{align}
The combined experimental new world average, based
on \citeres{Abi:2021gix,Bennett:2006fi}, was announced as
\begin{align}
\amu^{\rm exp} &= (11 659 206.1 \pm 4.1) \times 10^{-10}~.
\label{gmt-exp}
\end{align}
Compared with the SM prediction in \refeq{gmt-sm}, one arrives at a new
deviation of
\begin{align}
\Delta\amu &= (25.1 \pm 5.9) \times 10^{-10}~, 
\label{gmt-diff}
\end{align}
corresponding to a $4,2,\sig$ discrepancy.
We use this limit as a cut at the $\pm2\,\sig$ level.

\medskip
Recently a new lattice calculation for the leading order hadronic
vacuum polarization (LO HVP) contribution to
$\amu^{\rm SM}$~\cite{Borsanyi:2020mff} has been reported, which,
however, was not used in the new theory world
average, \refeq{gmt-sm}~\cite{Aoyama:2020ynm}. Consequently, we also do
not take this result into account, see also the discussions in
\citeres{CHS1,Lehner:2020crt,Borsanyi:2020mff,Crivellin:2020zul,Keshavarzi:2020bfy,deRafael:2020uif}. 
On the other hand, it is obvious that our conclusions would change
substantially if the result presented in \cite{Borsanyi:2020mff}
turned out to be correct. 

\medskip
In the MSSM the main contribution to \gmin2\ comes from one-loop
diagrams involving $\cha1-\Sn$ and $\neu1-\tilde \mu$ loops. 
In our analysis the MSSM contribution to \gmin2\
at two-loop order is calculated using {\tt GM2Calc}~\cite{Athron:2015rva},
implementing two-loop corrections
from \cite{vonWeitershausen:2010zr,Fargnoli:2013zia,Bach:2015doa}
(see also \cite{Heinemeyer:2003dq,Heinemeyer:2004yq}).

\bigskip
All other constraints are taken into account exactly as
in \citere{CHS1,CHS2,CHS3}. These are

\begin{itemize}

\item Vacuum stability constraints:\\
All points are checked to possess a stable and correct EW vacuum, e.g.\
avoiding charge and color breaking minima. This check is performed with
the public code {\tt Evade}~\cite{Hollik:2018wrr,Ferreira:2019iqb}.

\item Constraints from the LHC:\\
All relevant SUSY searches for EW particles are taken into account,
mostly via \CM~\cite{Drees:2013wra,Kim:2015wza,Dercks:2016npn} (see 
\citere{CHS1} for details on many analyses newly implemented by our group).
The LHC constraints that are most important for our scenarios come from
i) the production of $\cha1-\neu2$ pairs leading to three leptons and
$\met$ in the final state~\cite{Aaboud:2018jiw} ii) slepton-pair production leading to
two same flavour opposite sign leptons and $\met$ in the final state~\cite{Aad:2019vnb}.
Since all of our scenarios feature a low mass gap between the LSP
and the NLSP, the compressed spectra searches with the signature
of two soft leptons and $\met$ accompanied by an initial state radiation (ISR)
jet~\cite{Aad:2019qnd} also prove to be relevant in this case.
For the wino case, the disappearing track searches~\cite{Aaboud:2017mpt,Sirunyan:2020pjd}
are useful especially in the region of a low mass gap, $\Delta m \sim$ a
few hundred MeV.

\item
Dark matter relic density constraints:\\
For the experimental data we use the latest result from Planck~\cite{Planck},
either as a direct measurement, 
\begin{align}
\Omega_{\rm CDM} h^2 \; = \; 0.120,  \pm 0.001 \, , 
\label{OmegaCDM}
\end{align}
or as an upper bound, 
\begin{align}
\Omega_{\rm CDM} h^2 \; \le \; 0.120 \, . % \pm 0.001 \, .
\label{OmegaCDMlim}
\end{align}
The relic density in the MSSM is evaluated with
\MO~\cite{Belanger:2001fz,Belanger:2006is,Belanger:2007zz,Belanger:2013oya}.
In the latter case one needs an additional DM component
which would then bring the total DM density into agreement with the
Planck measurement in \refeq{OmegaCDM}. This could be, e.g., a SUSY
axion~\cite{Bae:2013bva}.

In the case of wino DM, because of the extremely small mass
splitting, the effect of ``Sommerfeld
enhancement''~\cite{Sommerfeld1931} can be very  
important. However, in \citere{CHS2} we argued why this does not have
any relevant effect on our analysis, and thus we do not take it into
account.

\item
Direct detection constraints of Dark matter:\\
We employ the constraint on the spin-independent
DM scattering cross-section $\ssi$ from
XENON-1T~\cite{XENON} experiment (which are always substantially more
relevant than the spin-dependent limits). 
The theoretical predictions are evaluated using the public code
\MO~\cite{Belanger:2001fz,Belanger:2006is,Belanger:2007zz,Belanger:2013oya}.
A combination with other DD experiments would put only very slightly
stronger limits. However, we will discuss the impact of possible
future limits and the neutrino floor below.

Here it should be noted that for
parameter points with $\Omega_{\tilde \chi} h^2 \; \le \; 0.118$
(i.e.\ $2\,\sig$ lower than the limit from Planck~\cite{Planck}, 
see \refeq{OmegaCDMlim}) we
rescale the cross-section with a factor of ($\Omega_{\tilde \chi} h^2$/0.118)
to take into account the fact that $\neu1$ provides only a fraction of the
total DM relic density of the universe.

\end{itemize}

Another potential set of constraints is given by the indirect
detection of DM. However, we do not impose these constraints
on our parameter space because of the well-known large uncertainties
associated with astrophysical factors like DM density profile as
well as theoretical corrections,
see~\citeres{Slatyer:2017sev,Hryczuk:2019nql,Rinchiuso:2020skh,Co:2021ion}.

%%%%%%%%%%%%%%%%%%%%%%%%%%%%%%%%%%%%%%%%%%%%%%%%%%%%%%%%%%%%%%%%%%%%%%%%%%
%%%%%%%%%%%%%%%%%%%%%%%%%%%%%%%%%%%%%%%%%%%%%%%%%%%%%%%%%%%%%%%%%%%%%%%%%%

\section{Parameter scan and analysis flow}
\label{sec:paraana}

\subsection{Parameter scan}
\label{sec:scan}

We scan the relevant MSSM parameter space to fully cover the allowed
regions of the relevant neutralino, chargino and slepton masses.
We follow the approach taken in \citeres{CHS1,CHS2,CHS3} and
investigate the five scenarios listed in \refse{sec:intro}. They are
given by the possible mass orderings of $M_1$, $M_2$, $\mu$ and $\mL$,
$\mR$. These masses determine the nature of the LSP and the NLSP, and
thus also the mechanism that reduces the relic DM density in the early
universe to or below the current value, see \refeqs{OmegaCDM},
(\ref{OmegaCDMlim}), i.e.\ coannihilation with the NLSP. We do not
take into account the possibility of pole annihilation, e.g.\ with the
$A$, the $h$ or the $Z$~boson. As argued in \citeres{CHS1,CHS2,CHS3}
these are rather remote possibilities in our set-up.
The five cases are covered as follows.
\begin{description}
\item
{\bf (A) Higgsino DM}\\
This scenario is characterized by a small value of~$\mu$ (as favored,
e.g., by naturalness
arguments~\cite{Baer:2012up,Baer:2013gva,Baer:2016lpj,Baer:2018rhs,Bae:2019dgg,Baer:2020vad})%
\footnote{
See \citere{Delgado:2020url} for 
a recent analysis in the higgsino DM scenario, requiring the LSP to
yield the full DM relic density.
}%
. Such a scenario is also naturally realized in Anomaly Mediation SUSY
breaking (AMSB, see e.g.\ \citere{Bagnaschi:2016xfg} and references therein).
We scan the following parameters: 
\begin{align}
  100 \gev \leq \mu \leq 1200 \gev \;,
  \quad 1.1 \mu \leq M_1, M_2 \leq 10 \mu\;, \notag \\
%  \quad 1.1  \mu \leq M_2 \leq 10 \mu, \;
  \quad 5 \leq \tb \leq 60, \; 
  \quad 100 \gev \leq \mL, \mR \leq 2000 \gev~.
\label{higgsino-dm}
\end{align}

\item
{\bf (B) Wino DM}\\
This scenario is characterized by a small value of $M_2$.
Also this type of scenario is naturally realized in the AMSB
(see e.g.\ \citere{Bagnaschi:2016xfg} and references therein).
We scan the following parameters: 
\begin{align}
  100 \gev \leq M_2 \leq 1500 \gev \;,
  \quad 1.1 M_2 \leq M_1, \mu \leq 10 M_2\;, \notag \\
%  \quad 1.1 M_2 \leq \mu \leq 10 M_2, \;
  \quad 5 \leq \tb \leq 60, \; 
  \quad 100 \gev \leq \mL, \mR \leq 2000 \gev~.
\label{wino-dm}
\end{align}
Here it should be noted that 
the choice of $M_2 \ll M_1, \mu$ at tree-level leads to an almost 
degenerate spectrum with $\mcha1 - \mneu1 = \order{1 {\rm\,eV}}$.
Going to the on-shell (OS) masses, yielding
a mass shift in $\mneu1$
and two other neutralino masses, the mass splitting between 
$\mcha1$ and $\mneu1$ is elevated which subsequently allows the decay
$\chapm1 \to \neu1 \pi^\pm$.
We refer to \citere{CHS2} for a detailed
description of our procedure.%
\footnote{The mass shift for our wino DM points has been calculated 
following \citeres{Fritzsche:2013fta,Heinemeyer:2017izw}.}

\item
{\bf (C) Mixed bino/wino DM}\\
Here we choose $M_1$ to be the smallest mass parameter and require
$\cha1$-coannihilation, given by a relatively small $M_2$.
The scan parameters are chosen as, 
\begin{align}
  100 \gev \leq M_1 \leq 1000 \gev \;,
  \quad M_1 \leq M_2 \leq 1.1 M_1\;, \notag \\
  \quad 1.1 M_1 \leq \mu \leq 10 M_1, \;
  \quad 5 \leq \tb \leq 60, \; \notag\\
  \quad 100 \gev \leq \mL \leq 1500 \gev, \; %\notag\\
  \quad \mR = \mL~.
\label{cha-coann}
\end{align}

\item
{\bf (D) Bino DM}\\
Also in this scenario we choose $M_1$ to be the smallest mass
parameter, but now require that a slepton is close in mass.
In this scenario ``accidentally'' the wino
component of the $\neu1$ can be non-negligible. However, this is not a
distinctive feature of this scenario.
We distinguish two cases: either the SU(2)
doublet sleptons, or the singlet sleptons are close in mass to the LSP.\\
{\bf (D1)} case-L: SU(2) doublet
\begin{align}
  100 \gev \leq M_1 \leq 1000 \gev \;,
  \quad M_1 \leq M_2 \leq 10 M_1 \;, \notag\\
  \quad 1.1 M_1 \leq \mu \leq 10 M_1, \;
  \quad 5 \leq \tb \leq 60, \; \notag\\
  \quad M_1 \leq \mL \leq 1.2 M_1, %\notag\\
  \quad M_1 \leq \mR \leq 10 M_1~. 
\label{slep-coann-doublet}
\end{align}

{\bf (D2)} case-R: SU(2) singlet
\begin{align}
  100 \gev \leq M_1 \leq 1000 \gev \;,
  \quad M_1 \leq M_2 \leq 10 M_1 \;, \notag \\
  \quad 1.1 M_1 \leq \mu \leq 10 M_1, \;
  \quad 5 \leq \tb \leq 60, \; \notag\\
  \quad M_1 \leq \mL \leq 10 M_1, \; 
  \quad M_1 \leq \mR \leq 1.2 M_1.
\label{slep-coann-singlet}
\end{align}
\end{description}
In all scans we choose flat priors of the parameter space and
generate \order{10^7} points. 

As discussed above, the mass parameters of the colored sector have
been set to high values, such that 
the resulting SUSY particle masses are outside the reach of the LHC,
and the light $\cp$-even Higgs-boson is in agreement with the LHC
measurements of the $\sim 125 \gev$ Higgs boson,
where the concrete values are not relevant for our analysis. Also $\MA$ has
been set to be above the TeV scale. 

%%%%%%%%%%%%%%%%%%%%%%%%%%%%%%%%%%%%%%%%%%%%%%%%%%%%%%%%%%%%%%%%%%%%%%%%%%%%%%%

\subsection{Analysis flow}
\label{sec:flow}

The data samples are generated by scanning randomly over the input parameter
range given above, where a flat prior has been taken for all parameters.
We use {\tt SuSpect}~\cite{Djouadi:2002ze}
as spectrum and SLHA file generator. In the next step the parameter
points are required to satisfy the $\chapm1$ mass limit from
LEP~\cite{lepsusy}. The SLHA output files as generated by
{\tt SuSpect} are then passed as input files to {\tt GM2Calc} and \MO~for
the calculation of \gmin2 and the DM observables, respectively. The parameter
points that satisfy the new \gmin2\ constraint, \refeq{gmt-diff}, the DM
relic density, \refeq{OmegaCDM} or (\ref{OmegaCDMlim}) (depending on
the scenario), the direct detection constraints (possibly with a
rescaled cross section) and the vacuum stability constraints, checked
with {\tt Evade}, are then taken to the final check 
against the LHC constraints as implemented in \CM.
The relevant branching ratios of the SUSY particles required by \CM\
are computed using {\tt SDECAY}~\cite{Muhlleitner:2003vg}.

%%%%%%%%%%%%%%%%%%%%%%%%%%%%%%%%%%%%%%%%%%%%%%%%%%%%%%%%%%%%%%%%%%%%%%%%%%

\section{Results}
\label{sec:results}

In this section we present our results for the DM implications in the
five scenarios. For each scenario we show the preferred ranges for the
LSP and NLSP masses, the DM relic density and the prospects for future
DD experiments.

%%%%%%%%%%%%%%%%%%%%%%%%%%%%%%%%%%%%%%%%%%%%%%%%%%%%%%%%%%%%%%%%%%%%%%%%%%

\subsection{Higgsino DM}
\label{sec:higgsino}

We start our discussion with the case of higgsino DM, as defined
in \refse{sec:scan}. 
The plots show only points that are in agreement with all theoretical
and experimental constraints. 

%%%%%%%%%%%%%%%%%%%%%%%%%%%% F I G U R E %%%%%%%%%%%%%%%%%%%%%%%%%%%%%%
\begin{figure}[htb!]
\vspace{3em}
\centering
\includegraphics[width=0.72\textwidth]{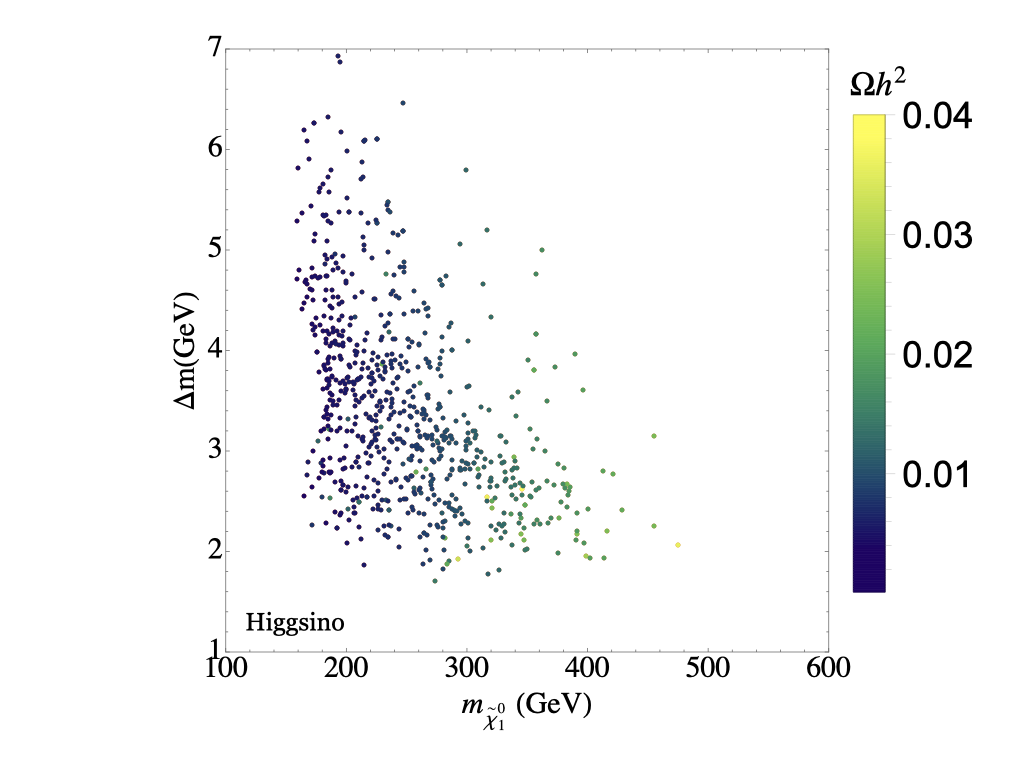}\\
\includegraphics[width=0.82\textwidth]{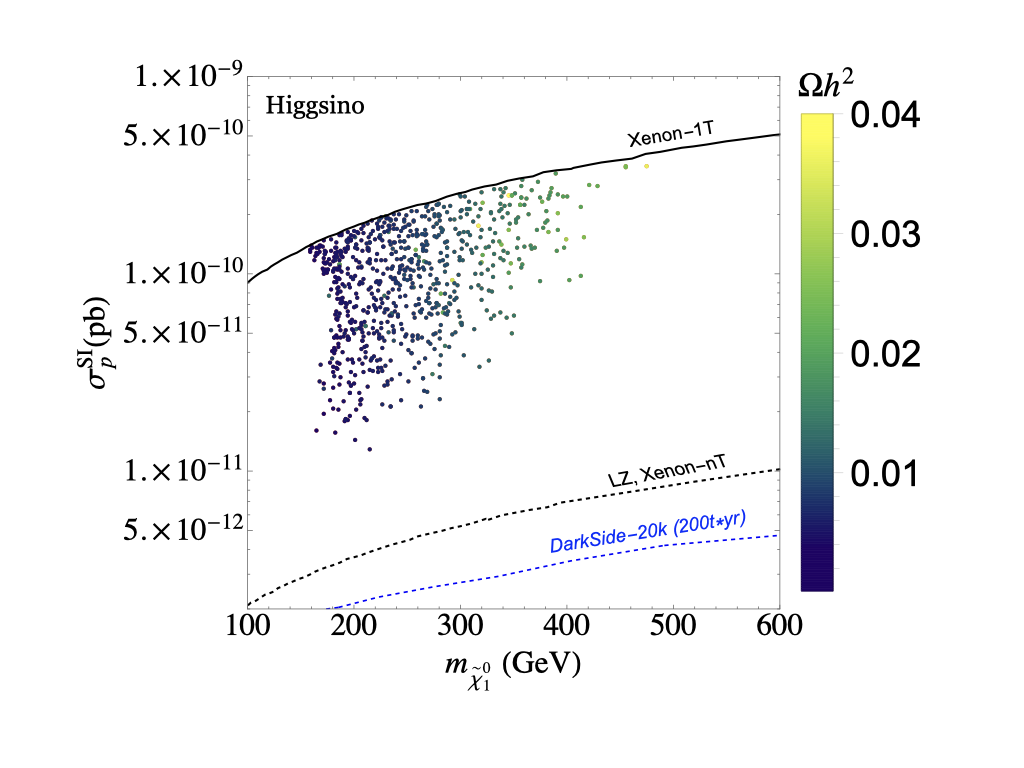}
\caption{The results of our parameter scan in the higgsino DM
scenario.
Upper plot: $\mneu1$--$\De m$ plane ($\De m = \mcha1 - \mneu1$).
Lower plot: $\mneu1$--$\ssi$ plane.
The color code indicates the DM relic density.
}
\label{fig:higgsino}
\end{figure}
%%%%%%%%%%%%%%%%%%%%%%%%%%%% F I G U R E %%%%%%%%%%%%%%%%%%%%%%%%%%%%%%

In \reffi{fig:higgsino} we show the results of our parameter
scan. The upper plot shows the $\mneu1$--$\De m$ plane, with
$\De m := \mcha1 - \mneu1$, and $\mneu2 \approx \mcha1$. 
The allowed LSP masses range from $\sim 150 \gev$ to about
$\sim 500 \gev$, where $\De m$ is found in the range between
$\sim 1.5 \gev$ and $\sim 7 \gev$. Larger DM masses are reached for
smaller mass gaps. 
The color code indicates the relic density. Low LSP masses correspond
to the lowest density, below $\Omega_{\neu1}h^2 \lsim 0.01$, going up
to $\sim 0.04$ for the largest values of $\mneu1$. The full relic
density would be reached for $\mneu1 \sim 1 \tev$. This, however,
would be in disagreement with the \gmin2\ prediction, and
consequently, only lower densities are found.

We now turn to the prediction for the direct detection of DM in the
higgsino scenario. Here it is important to note that
the dominant contribution to DM scattering comes
from the exchange of a light $\cp$-even Higgs
boson in the $t$-channel. The corresponding $h\neu1\neu1$ coupling
is given at tree level by~\cite{Hisano:2004pv}
\begin{eqnarray}
c_{h\neu1\neu1}&\simeq&
- \frac12 (1 + \sin2\beta)
\LP \tan^2\theta_\mathrm{w} \frac{\MW}{M_1-\mu}
+ \frac{\MW}{M_2-\mu} \RP\,,
\label{ddhiggsino}
\end{eqnarray}
where $\mu > 0$ has been assumed (as given in our scan). One can see
that the coupling becomes large for $\mu \sim M_2$ or $\mu \sim M_1$.
Consequently, the XENON-1T DD bound pushes the allowed parameter space into
the almost pure higgsino-LSP region, with negligible bino and wino
component, i.e.\ to larger values for $M_2/\mu$ and $M_1/\mu$. 
In the lower plot of \reffi{fig:higgsino} we show the prediction
for the direct detection prospects in the higgsino DM scenario.
The allowed points are displayed
in the $\mneu1$--$\ssi$ plane, where again the color code indicates
the DM relic density. Here it should be remembered that we
rescale the cross-section with a factor of ($\Omega_{\tilde \chi} h^2$/0.118)
to take into account the fact that $\neu1$ provides only a fraction of the
total DM relic density of the universe.
The points are by construction bounded from
above by the XENON-1T limit~\cite{XENON}. We also show the projection for
the exclusion reach of XENON-nT~\cite{Aprile:2020vtw} and of the LZ
experiment~\cite{LZ} as black dashed line (which effectively agree with each
other). Furthermore, we show the projection of the
DarkSide~\cite{DarkSide} experiment,
which can go down to even lower cross sections, as blue dashed line.
One can see that the full parameter space will be covered already by
XENON-nT and/or LZ. Also DarkSide with its lower reach will cover the
complete higgsino DM scenario.

%%%%%%%%%%%%%%%%%%%%%%%%%%%%%%%%%%%%%%%%%%%%%%%%%%%%%%%%%%%%%%%%%%%%%%%%%%

\subsection{Wino DM}
\label{sec:wino}

The next case we present here is the wino DM case, as discussed
in \refse{sec:scan}. 
As before, the plots show only points that are in agreement with all
theoretical and experimental constraints. 

\begin{figure}[htb!]
	\vspace{2em}
\centering
\includegraphics[width=0.77\textwidth]{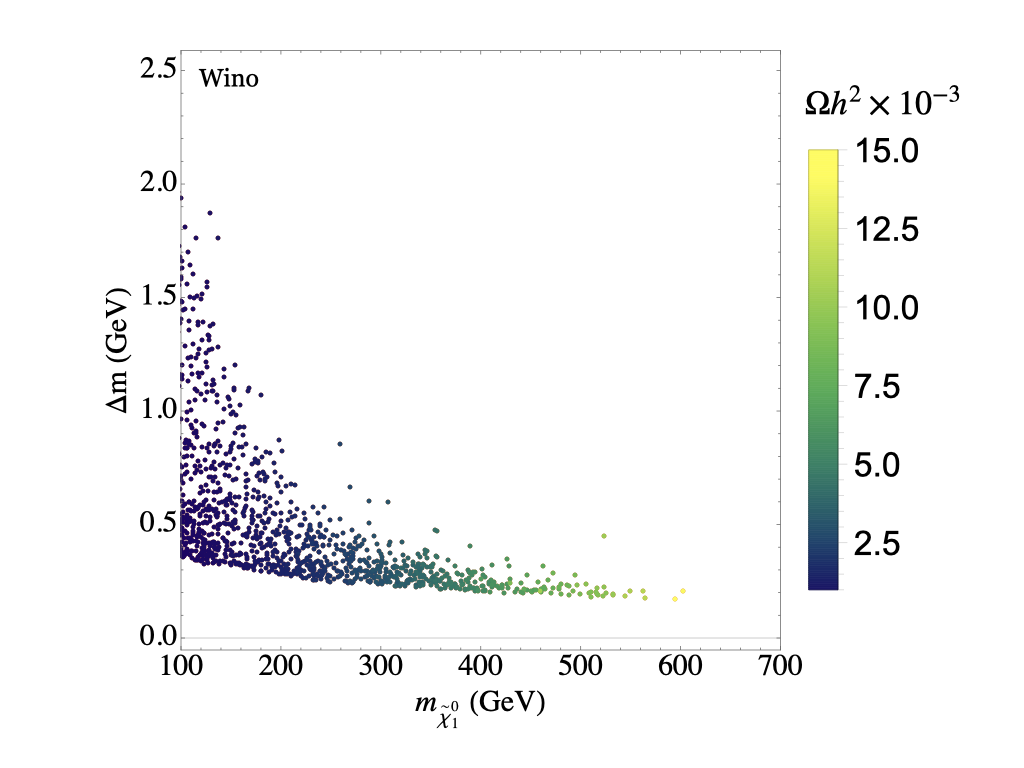}\\
\mbox{}\hspace{-5mm}\includegraphics[width=0.80\textwidth]{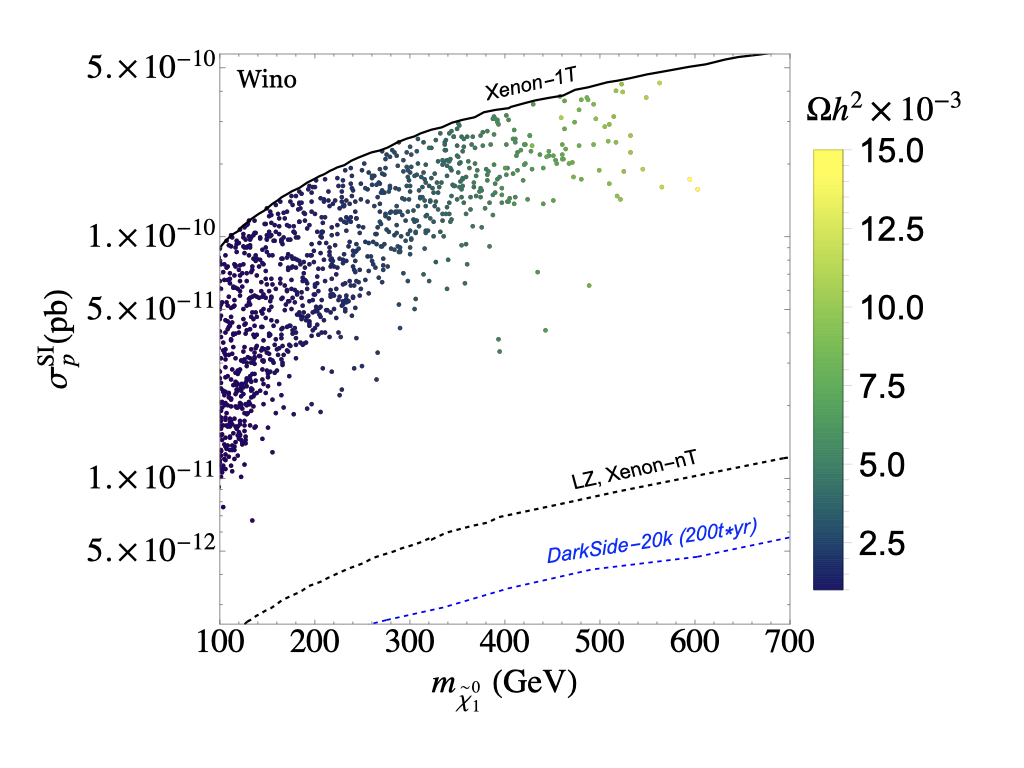}
\caption{The results of our parameter scan in the wino DM
scenario.
Upper plot: $\mneu1$--$\De m$ plane ($\De m = \mcha1 - \mneu1$).
Lower plot: $\mneu1$--$\ssi$ plane.
The color code indicates the DM relic density.
}
\label{fig:wino}
\end{figure}

In \reffi{fig:wino} we show the results of our parameter
scan. The upper plot shows the $\mneu1$--$\De m$ plane, with
$\De m := \mcha1 - \mneu1$.
The allowed LSP masses range from $\sim 100 \gev$, where we started
our scan, to about 
$\sim 600 \gev$, where $\De m$ is found in the range between
$\sim 0.2 \gev$ and $\sim 2 \gev$.
Here it should be remembered that
the choice of $M_2 \ll M_1, \mu$ leads to an approximately
degenerate spectrum at tree-level with $\mcha1 - \mneu1 = \order{1 {\rm\,eV}}$.
Only by going to OS masses,
yielding a mass shift in $\mneu1$
and two other neutralino masses and hence with the raised splitting between 
$\mcha1$ and $\mneu1$, the decay
$\chapm1 \to \neu1 \pi^\pm$ is allowed. The disappearing track searches at
the LHC~\cite{Sirunyan:2020pjd} then cut away the smallest $\De m$
region (see \citere{CHS2} for details), 
resulting in the lower limit displayed in \reffi{fig:wino}. 
Larger DM masses are reached for smaller mass gaps.
The color code indicates the relic density. Low LSP masses correspond
to the lowest density, below $\Omega_{\neu1}h^2 \lsim 0.0025$ (i.e.\
even smaller by a factor of four compared to the higgsino case), going up
to $\sim 0.015$ for the largest values of $\mneu1$. The full relic
density would be reached for $\mneu1 \sim 3 \tev$. However, as in the
higgsino case, this would be in disagreement with the \gmin2\
prediction, and consequently, only substantially lower densities are found.

%%%%%%%%%%%%%%%%%%%%%%%%%%%% F I G U R E %%%%%%%%%%%%%%%%%%%%%%%%%%%%%%
%%%%%%%%%%%%%%%%%%%%%%%%%%%% F I G U R E %%%%%%%%%%%%%%%%%%%%%%%%%%%%%%

The overall allowed parameter space is furthermore bounded
from ``above'' by the  DD limits, which cut away larger mass differences,
which can be understood as follows. For a wino-like $\neu1$, the
$h\neu1\neu1$ coupling is given by~\cite{Hisano:2004pv}
\begin{eqnarray}
c_{h\neu1\neu1}\simeq
\frac{\MW}{M_2^2-\mu^2}(M_2+\mu\sin2\beta)\,.
\label{ddwino}
\end{eqnarray}
In the limit of $||\mu|-M_2|\gg \MZ$ and assuming also that the
$h$-exchange dominates over the $H$~contribution in the (spin
independent) DD bounds (i.e.\ the $\cp$-odd Higgs also does not
contribute),
the $h\neu1\neu1$ coupling becomes large at $\mu \sim M_2$.
The tree level mass splitting between the two wino-like states
$\chapm1$ and $\neu1$ (generated mainly by the mixing of the lighter
chargino with the charged higgsino) is given by~\cite{Ibe:2012sx}
\begin{eqnarray}
\De m (= \mcha1 - \mneu1) \simeq
\frac{ \MW^4 (\sin 2\beta)^2\tan^2 \theta_{\mathrm{w}} }{ (M_1 - M_2) \mu^2 },
\label{delmtree}
\end{eqnarray}
assuming $|M_1 - M_2| \gg \MZ$. Therefore, 
the mass splitting increases for
smaller $\mu$ values with a simultaneous increase in DD cross-section.

In the lower plot of \reffi{fig:wino} we show the prediction
for the direct detection prospects in the wino DM scenario.
The allowed points are displayed
in the $\mneu1$--$\ssi$ plane, where again the color code indicates
the DM relic density. As in the higgsino DM case we
re-scale the cross-section with a factor of ($\Omega_{\tilde \chi} h^2$/0.118)
to take into account the fact that $\neu1$ provides only a fraction of the
total DM relic density of the universe.
By construction the points are bounded from above by the XENON-1T
limit~\cite{XENON}, where the smallest $\mu/M_2$ values are found. 
As discussed above, the lower limit is given by the disappearing track
searches at the LHC~\cite{Sirunyan:2020pjd}, i.e.\ small mass
splittings. Also in this plot we show as black dashed line the projected
limit of XENON-nT/LZ and as blue dashed
line the one of DarkSide.
One can see that the XENON-nT and/or LZ result will
either firmly exclude or detect a wino DM candidate, possibly in
conjunction with improved disappearing track searches at the LHC.
The same holds for the DarkSide experiment.

%%%%%%%%%%%%%%%%%%%%%%%%%%%%%%%%%%%%%%%%%%%%%%%%%%%%%%%%%%%%%%%%%%%%%%%%%%

\subsection{Bino/wino DM with \boldmath{$\cha1$}-coannihilation}
\label{sec:chaco}

In this section we analyze the case of bino/wino DM, as defined
in \refse{sec:scan}. In this scenario $\cha1$-coannihilation is
responsible for finding the DM relic density either in full agreement
with the Planck measurements, see \refeq{OmegaCDM}, or is found to
be smaller, see \refeq{OmegaCDMlim}. 

%%%%%%%%%%%%%%%%%%%%%%%%%%%% F I G U R E %%%%%%%%%%%%%%%%%%%%%%%%%%%%%%
\begin{figure}[htb!]
%	\vspace{4em}
\centering
\includegraphics[width=0.77\textwidth]{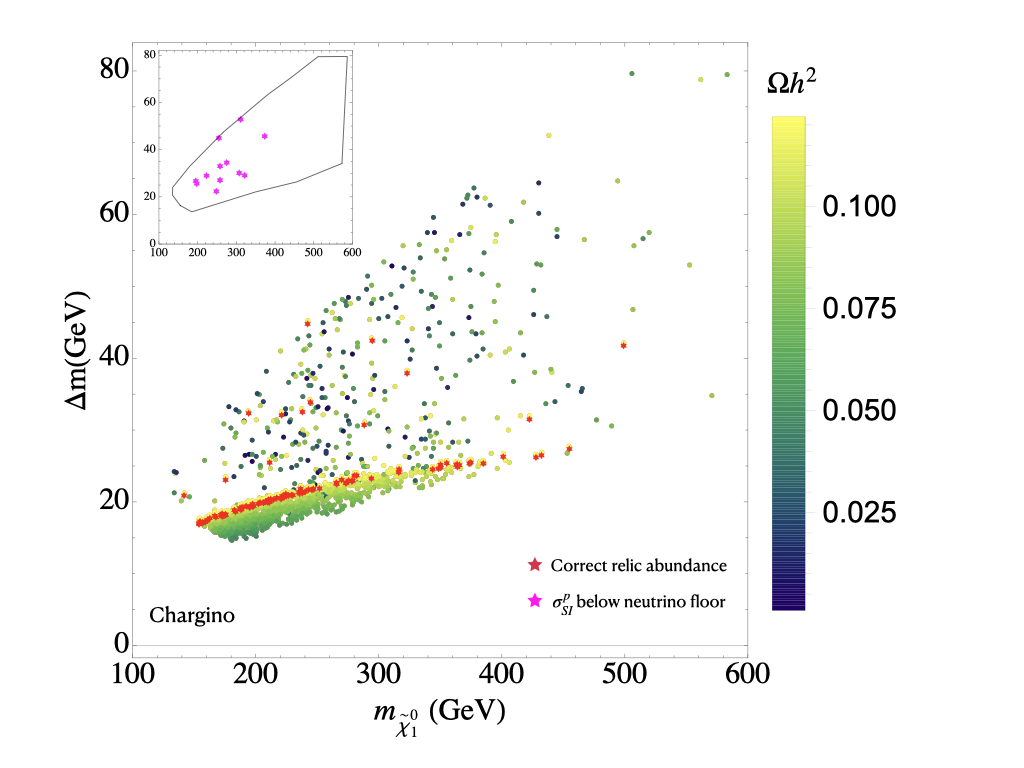}\\
\includegraphics[width=0.77\textwidth]{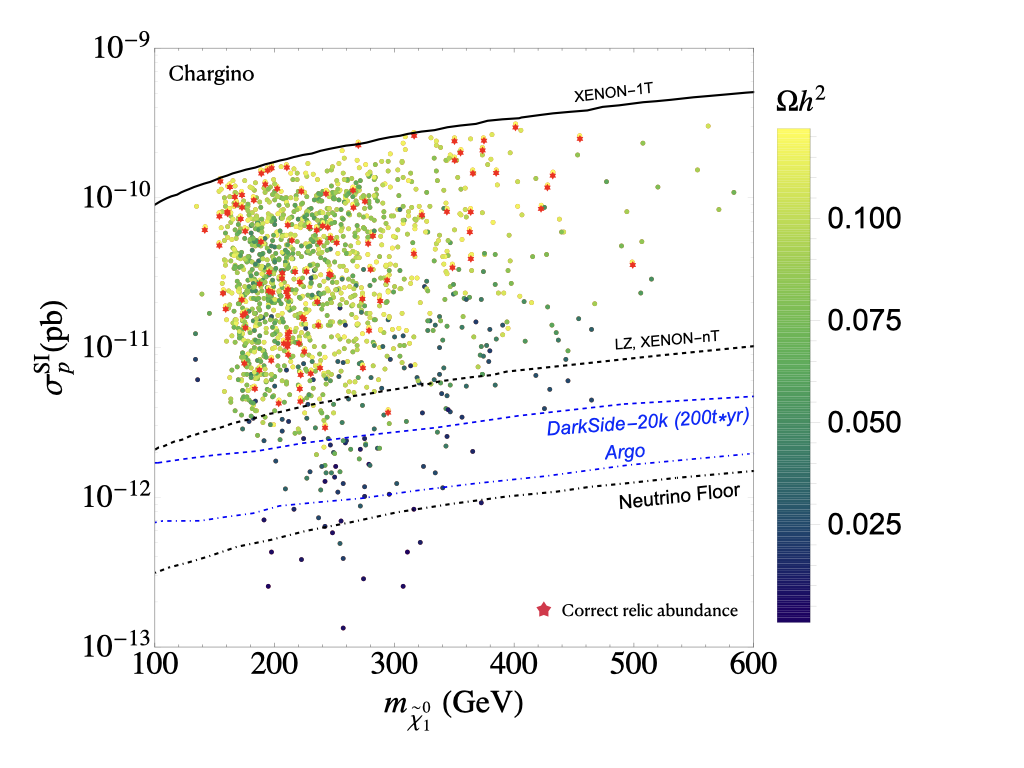}
\caption{The results of our parameter scan in the bino/wino DM
scenario with $\cha1$-coannihilation.
Upper plot: $\mneu1$--$\De m$ plane ($\De m = \mcha1 - \mneu1$).
Lower plot: $\mneu1$--$\ssi$ plane.
The color code indicates the DM relic density. Red points are in full
agreement with the Planck measurement.
The magenta points shown in the inlay in the upper plot indicate the
points below the neutrino floor, where the solid line indicates the
overall allowed parameter space. 
}
\label{fig:chaco}
\end{figure}
%%%%%%%%%%%%%%%%%%%%%%%%%%%% F I G U R E %%%%%%%%%%%%%%%%%%%%%%%%%%%%%%

In the upper plot of \reffi{fig:chaco} we show our results in the
$\mneu1$--$\De m$ plane (with $\De m = \mcha1 - \mneu1$).
The color coding indicates
the DM relic density, where the red points correspond to full agreement
with the Planck measurement, see \refeq{OmegaCDM}.
The magenta points shown in the inlay are found below the neutrino
floor, see the discussion below. The solid line surrounding the points
indicates the overall allowed parameter space in this plane.
By definition of $\cha1$-coannihilation, the points are found for
relatively low values of $\De m$, between $\sim 10 \gev$ and
$\sim 60 \gev$.
Two ``populations'' can be observed. One large group of parameter points
are found at $\De m \sim 20 \gev$. In these points only the chargino
contributes relevantly to the correct relic abundance. For the 
sparsely distributed region in the higher $\De m$,
mostly sleptons contribute to the coannihilation.
Concerning the ``pure'' $\cha1$-coannihilation points, due to the small
mass splitting, it will be more complicated to detect these points at the
(HL-)LHC, see also the discussion in \refse{sec:future-pp}.

The prediction for the DD experiments is demonstrated 
in the lower plot of \reffi{fig:chaco}. We show the $\mneu1$--$\ssi$
plane, again with the color coding indicating the DM relic density.
As in the previous cases, for the points with a lower relic density we
rescale the cross-section with a factor of ($\Omega_{\tilde \chi} h^2$/0.118)
to take into account the fact that $\neu1$ provides only a fraction of the
total DM relic density of the universe.
By construction, the upper limit of the points is provided by the
XENON-1T limit. In addition to the XENON-nT, LZ and DarkSide limits we
also show the anticipated reach of the Argo experiment~\cite{Argo}, as
well as the neutrino floor~\cite{neutrinofloor}.
It is evident that for the lowest DM relic density,
the DD cross-section is considerably scaled down.
The red points, i.e.\ the ones with correct relic abundance,
spread out even slightly below the future XENON-nT/LZ
limit, but all lie above the anticipated reach of DarkSide and Argo.
Cross sections lower than the DarkSide reach are only found for low values of
$\Omega_{\tilde\chi} h^2$. Those points can reach even values below the
neutrino floor. As can be seen in the inlay in the upper plot
of \reffi{fig:chaco}, 
these points (indicated by magenta stars) do not exceed $\mneu1 \sim 400 \gev$.
On the other hand, no clear pattern w.r.t. $\De m$ can be observed for
these points. In \refse{sec:complementarity} we will discuss the
complementarity of the direct detection experiments with the anticipated
reach at the (HL-)LHC and possible future $e^+e^-$ collider
experiments. From the upper limit on the masses of the points below the
neutrino floor, it becomes apparent already that one could
cover them at an $e^+e^-$ collider with $\sqrt{s} \lsim 1 \tev$ via
$e^+e^- \to \neu1\neu1\ga$. This demonstrates the complementarity of
DD experiments and future (linear) $e^+e^-$ colliders.

%%%%%%%%%%%%%%%%%%%%%%%%%%%%%%%%%%%%%%%%%%%%%%%%%%%%%%%%%%%%%%%%%%%%%%%%%%

\subsection{Bino DM with \boldmath{$\Slpm$}-coannihilation case-L}
\label{sec:caseL}

%%%%%%%%%%%%%%%%%%%%%%%%%%%% F I G U R E %%%%%%%%%%%%%%%%%%%%%%%%%%%%%%
\begin{figure}[htb!]
	\vspace{2em}
\centering
\includegraphics[width=0.73\textwidth]{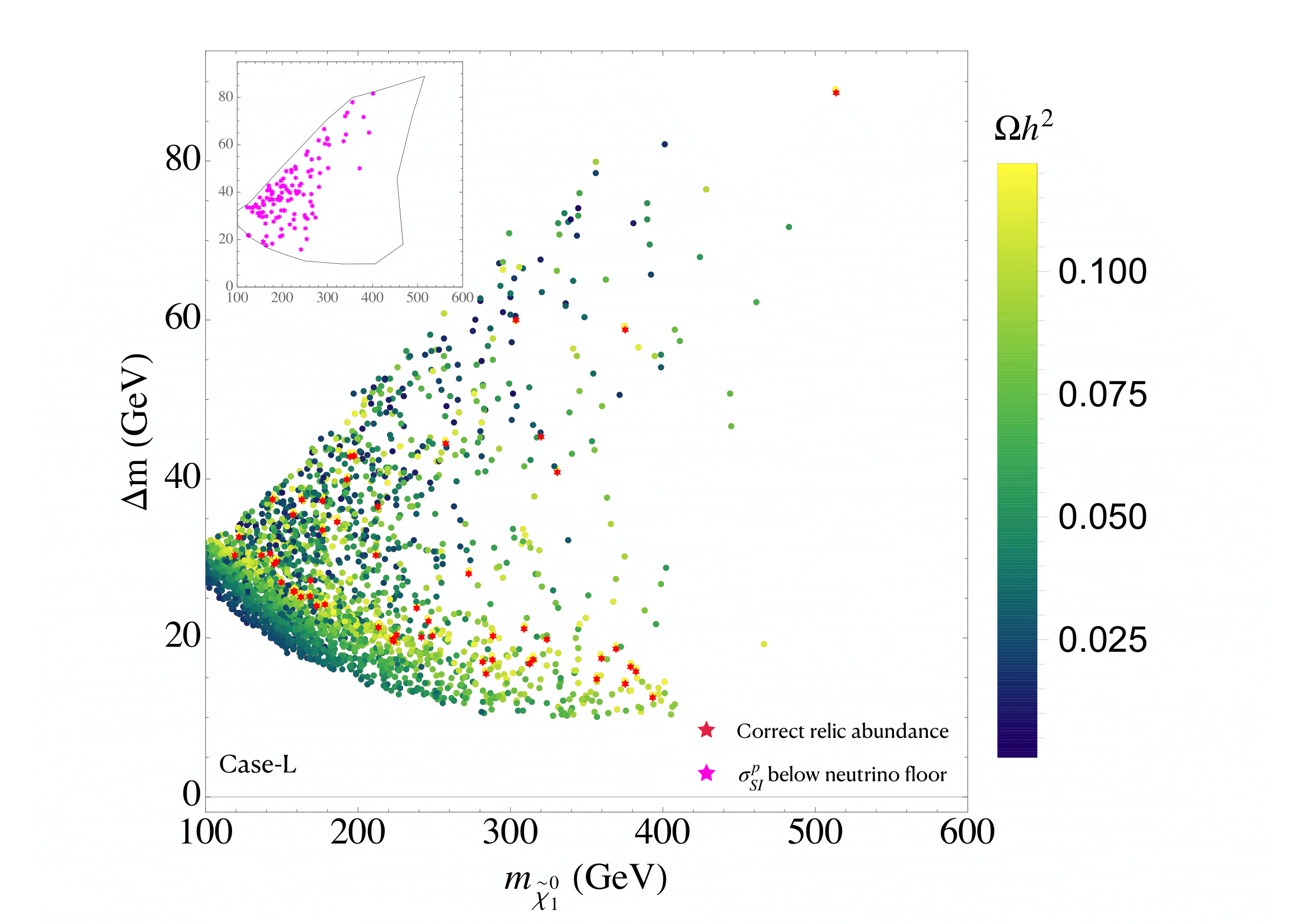}\\[1em]
\includegraphics[width=0.73\textwidth]{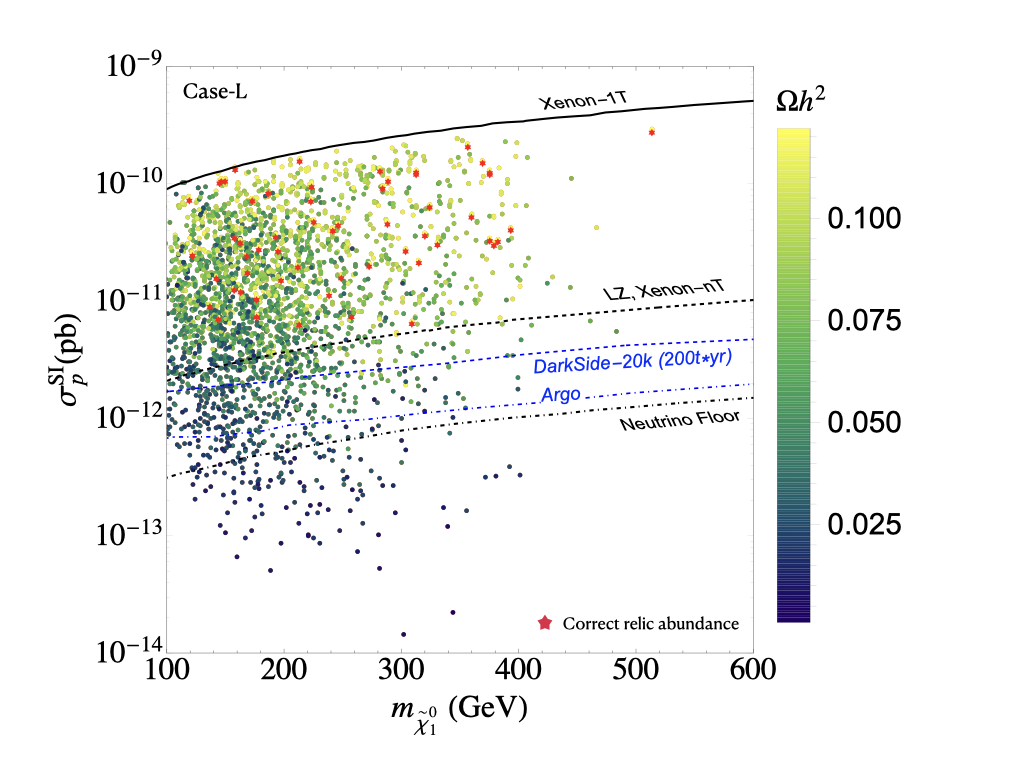}
\caption{The results of our parameter scan in the bino DM
scenario with $\Slpm$-coannihilation case-L. 
Upper plot: $\mneu1$--$\De m$ plane ($\De m = \msmu1 - \mneu1$).
Lower plot: $\mneu1$--$\ssi$ plane.
The color code indicates the DM relic density. Red points are in full
agreement with the Planck measurement.
The magenta points shown in the inlay in the upper plot indicate the
points below the neutrino floor, where the solid line indicates the
overall allowed parameter space.
}
\label{fig:caseL}
\end{figure}
%%%%%%%%%%%%%%%%%%%%%%%%%%%% F I G U R E %%%%%%%%%%%%%%%%%%%%%%%%%%%%%%

%\clearpage
We now turn to the case of bino DM with $\Slpm$-coannihilation. As
discussed in \refse{sec:scan} we distinguish two cases, depending
which of the two slepton soft SUSY-breaking parameters is set to be close to
$\mneu1$. We start with the case-L, where we chose $\msl{L} \sim M_1$,
i.e.\ the left-handed 
charged sleptons as well as the sneutrinos are close in mass to the LSP.
As analyzed in \citeres{CHS1,CHS2}, one finds
that all six sleptons are close in mass and differ by less
than $\sim 50 \gev$.

In the upper plot of \reffi{fig:caseL} we show the results of our scan
in the $\mneu1$--$\De m$ plane (with $\De m =\msmu1 - \mneu1$).
The color coding indicates the DM relic
density, where the red points correspond to full agreement with the
Planck measurement, see \refeq{OmegaCDM}.
The magenta points shown in the inlay are found below the neutrino
floor, see below, where the solid line indicates the
overall allowed parameter space.
By definition of $\Slpm$-coannihilation the points are found for
relatively low $\De m$, with mass differences between $\sim 10 \gev$
and $\sim 80 \gev$. For each $\mneu1$ the smallest achievable $\De m$
values result in an underabundance of DM. Apart from that, no
clear pattern can be observed for the location of the red points
(fulfilling exactly the Planck measurements).
Concerning the magenta points (below the neutrino floor) they are
found only for $\mneu1 \lsim 400 \gev$, i.e.\ making them potentially
easier to access at future collider experiments.
In \refse{sec:future-pp} we will discuss in more detail how
the various population of points may be tested at the (HL-)LHC or a
future $e^+e^-$ collider.

The prediction for the DD experiments is presented
in the lower plot of \reffi{fig:caseL}. We show the $\mneu1$--$\ssi$
plane, again with the color coding indicating the DM relic density.
As in the previous cases, for the points with a lower relic density we
rescale the cross-section with a factor of ($\Omega_{\tilde \chi} h^2$/0.118)
to take into account the fact that $\neu1$ provides only a fraction of the
total DM relic density of the universe.
By construction, the upper limit of the points is provided by the
XENON-1T limit. 
The red points, i.e.\ the ones in full agreement with the Planck
measurement, are all above the future XENON-nT/LZ limit, i.e.\ the can
all be tested in future DD experiments. This also holds for DarkSide
(blue dashed) and Argo (blue dot-dashed), which have an even higher
anticipated sensitivity. 
However, going to lower relic densities, on can observe that very low cross
sections are reached for the lowest values of $\Omega_{\tilde\chi} h^2$.
Those points can reach even values substantially below the
neutrino floor, i.e.\ the prospects to cover them in DD experiments are
currently unclear. On the other hand, as can be seen in the upper plot
of \reffi{fig:caseL}, 
these points (indicated by magenta stars in the inlay) do not exceed
$\mneu1 \sim 500 \gev$. Their discovery prospects at the HL-LHC and
future $e^+e^-$ colliders with $\sqrt{s} = 1000 \gev$, i.e.\ the
complementarity of DD and collider experiments will be discussed
in \refse{sec:complementarity}.

%%%%%%%%%%%%%%%%%%%%%%%%%%%%%%%%%%%%%%%%%%%%%%%%%%%%%%%%%%%%%%%%%%%%%%%%%%%%%%

\subsection{Bino DM with \boldmath{$\Slpm$}-coannihilation case-R}
\label{sec:caseR}

We now turn to our fifth scenario, bino DM with $\Slpm$-coannihilation case-R,
where in the scan we require the ``right-handed'' sleptons to be close
in mass with the LSP. Here it should be kept in mind that in our notation
we do not mass-order the sleptons: for negligible mixing as it is
given for selectrons and smuons the ``left-handed'' (``right-handed'')
slepton corresponds to $\Sl_1$ ($\Sl_2$). As discussed
in \citeres{CHS1,CHS2}, in this scenario all
relevant mass scales are required to be relatively light by the
\gmin2\ constraint.

%%%%%%%%%%%%%%%%%%%%%%%%%%%% F I G U R E %%%%%%%%%%%%%%%%%%%%%%%%%%%%%%
\begin{figure}[htb!]
%	\vspace{2em}
\centering
\includegraphics[width=0.75\textwidth]{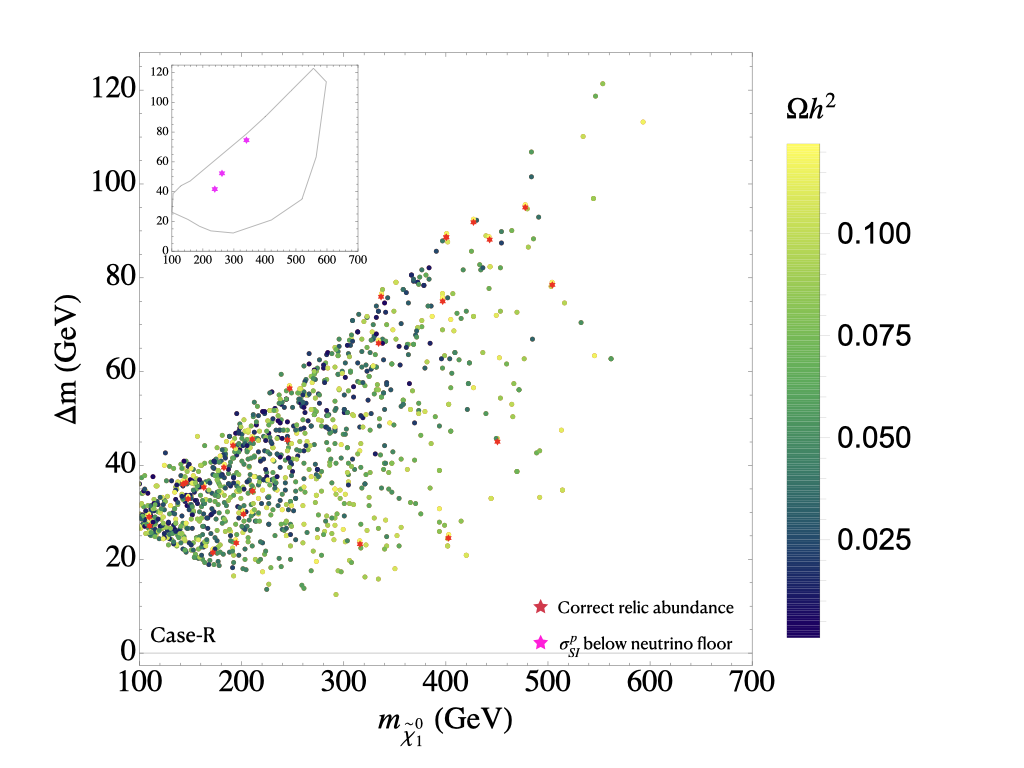}\\[1em]
\includegraphics[width=0.75\textwidth]{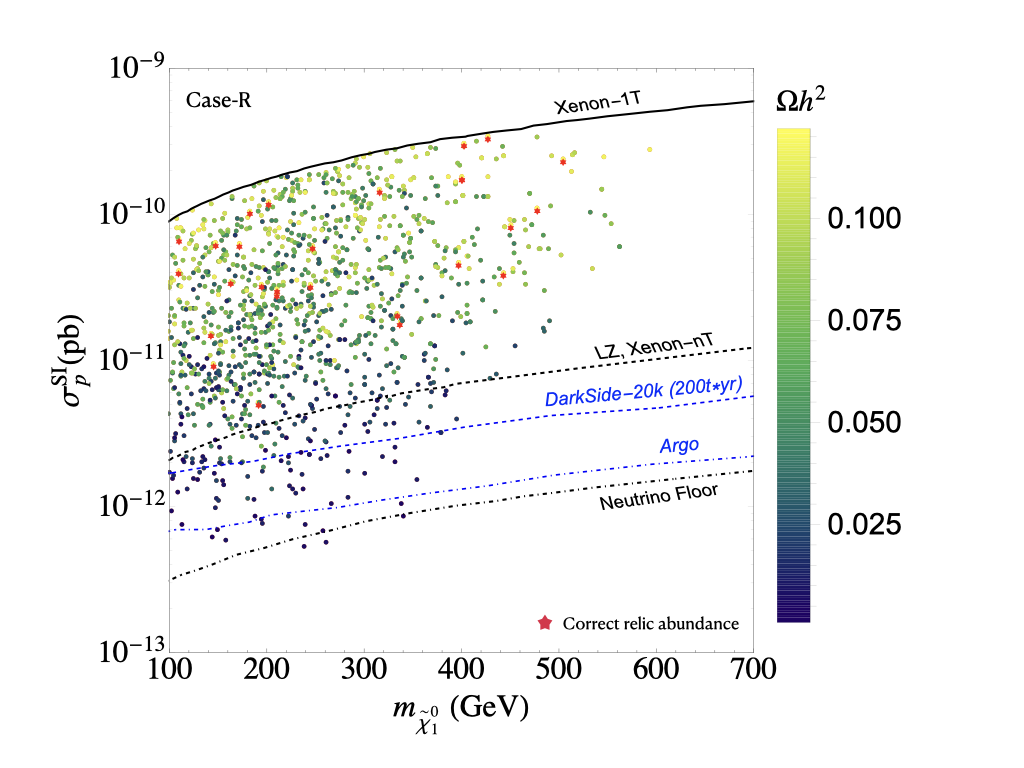}
\caption{The results of our parameter scan in the bino DM
scenario with $\Slpm$-coannihilation case-R. 
Upper plot: $\mneu1$--$\De m$ ($\De m = \msmu2 - \mneu1$).
Lower plot: $\mneu1$--$\ssi$ plane.
The color code indicates the DM relic density. Red points are in full
agreement with the Planck measurement.
The magenta points shown in the inlay in the upper plot indicate the
points below the neutrino floor, where the solid line indicates the
overall allowed parameter space.
}
\label{fig:caseR}
\end{figure}
%%%%%%%%%%%%%%%%%%%%%%%%%%%% F I G U R E %%%%%%%%%%%%%%%%%%%%%%%%%%%%%%

In the upper plot of \reffi{fig:caseR} we show the results of our scan
in the $\mneu1$--$\De m$ plane (with $\De m = \msmu2 - \mneu1$).
The color coding indicates the DM relic 
density, where the red points correspond to full agreement with the
Planck measurement, see \refeq{OmegaCDM}.
The (three) magenta points shown in the inlay are found below the neutrino
floor, see below, where the solid line indicates the
overall allowed parameter space.
By definition of $\Slpm$-coannihilation the points are found for
relatively low $\De m$.
Contrary to case-L the red points (fulfilling exactly the Planck
measurements) are distributed over the whole allowed parameter space.
The sparse magenta points (below the neutrino floor, see below) are
found between $\sim 200 \gev \lsim \mneu1 \lsim 350 \gev$.
In \refse{sec:future-pp} we will discuss in more detail how
the various population of points may be tested at the (HL-)LHC or a
future $e^+e^-$ collider. 

The prediction for the DD experiments is presented
in the lower plot of \reffi{fig:caseR}. We show the $\mneu1$--$\ssi$
plane, again with the color coding indicating the DM relic density.
As before, for the points with a lower relic density we
rescale the cross-section with a factor of ($\Omega_{\tilde \chi} h^2$/0.118)
to take into account the fact that $\neu1$ provides only a fraction of the
total DM relic density of the universe.
By construction, the upper limit of the points is provided by the
XENON-1T limit. 
The red points, i.e.\ the ones in full agreement with the Planck
measurement, spread out substantially below the current XENON-1T 
limit. 
However, they do not go below the future anticipated XENON-nT/LZ limit
(black dashed line), i.e.\ they can be covered by future DD experiments.
This also holds for DarkSide
(blue dashed) and Argo (blue dot-dashed), which have an even higher
anticipated sensitivity. 

Going to lower relic densities, one can observe that, as in the
previously analyzed cases, very low cross
sections are reached for the lowest values of $\Omega_{\tilde\chi} h^2$.
Those points can reach even values going down to the 
neutrino floor, with three of them even below.
It should be noted here that these three points (marked as magenta stars
in the upper inlay) are just on the border of the neutrino floor, which
may be subject to some uncertainties~\cite{Boehm:2018sux}.
Consequently, no firm conclusion can be drawn for them.
On the other hand, as can be seen in the upper plot
of \reffi{fig:caseR}, all points below the XENON-nT/LZ limit do not exceed
$\mneu1 \sim 400 \gev$. This leads to possibly very good prospects for
their discovery at the HL-LHC or a 
future $e^+e^-$ colliders with $\sqrt{s} = 1000 \gev$.
The corresponding complementarity of DD and collider experiments will be
discussed in the next section.

%%%%%%%%%%%%%%%%%%%%%%%%%%%%%%%%%%%%%%%%%%%%%%%%%%%%%%%%%%%%%%%%%%%%%%%%%%
%%%%%%%%%%%%%%%%%%%%%%%%%%%%%%%%%%%%%%%%%%%%%%%%%%%%%%%%%%%%%%%%%%%%%%%%%%

\section{Complementarity with future collider experiments}
\label{sec:complementarity}

In this section we analyze the complementarity between future DD
experiments and searches at colliders. We concentrate on the parameter
points that are below the anticipated limits of XENON-nT and LZ, and in
particular on the points below the neutrino floor. 
We first show the prospects for searches for EW SUSY particles at the
approved HL-LHC~\cite{CidVidal:2018eel} and then at possible future
high-energy $e^+e^-$ colliders, such as the ILC~\cite{ILC-TDR,LCreport}
or CLIC~\cite{CLIC,LCreport}.

%%%%%%%%%%%%%%%%%%%%%%%%%%%%%%%%%%%%%%%%%%%%%%%%%%%%%%%%%%%%%%%%%%%%%%%%%%

\subsection{HL-LHC prospects}
\label{sec:future-pp}

The prospects for BSM phenomenology at the HL-LHC have been summarized
in \citere{CidVidal:2018eel} for a 14~TeV run with 3 \iab\ of
integrated luminosity per detector.
For the wino, higgsino and bino/wino with $\cha1$-coannihilation DM scenarios,
the most relevant constraints may be derived either by searches specially
designed to look for compressed spectra with low mass-splitting
between $\cha1, \neu2$ and $\neu1$, or complementary by searching
for slepton pair-production at the HL-LHC.
The projected discovery and 95\% confidence level (C.L.)
exclusion regions for the former search have been published by both
CMS and ATLAS collaborations for the higgsino simplified model scenario.
A naive application of the projected exclusion contours on our
model parameter space (i.e.\ not taking into account the variation due
to the difference in production cross section)
shows that the higgsino and bino/wino scenarios will be covered in part
by the HL-LHC, see, e.g., Fig.~22 in \citere{CHS2}.
However, for the wino scenario the mass-splitting is too low
to be probed by the compressed spectra searches. However, in this
case, the improved HL-LHC sensitivity to disappearing track searches can
prove to be useful, particularly in the region of very low
mass-splittings~\cite{CHS2}.For $\Delta m \sim 170 \mev$, the
HL-LHC can probe wino masses upto about $900 \gev$ and $500 \gev$ at the
95\% C.L., for the optimistic and conservative background estimations
respectively. 

\begin{figure}[htb!]
%	\vspace{4em}
\centering
\includegraphics[width=0.45\textwidth]{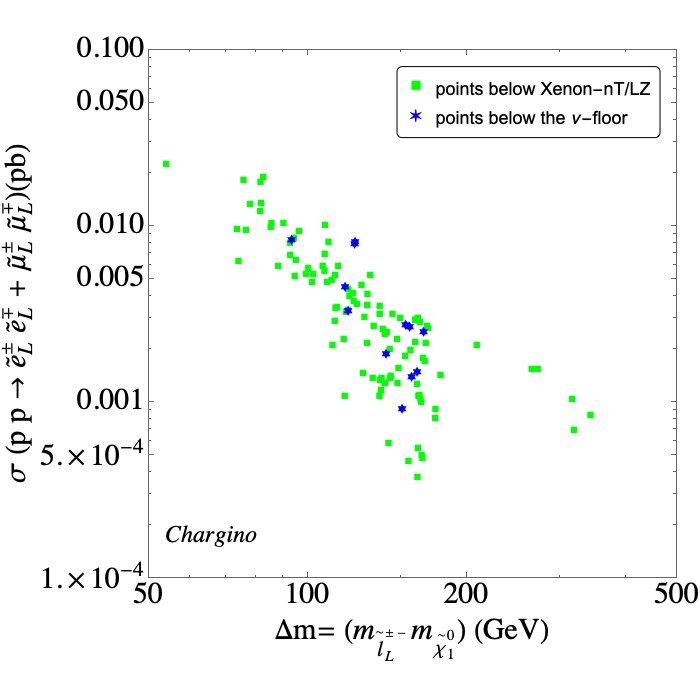}
\includegraphics[width=0.45\textwidth]{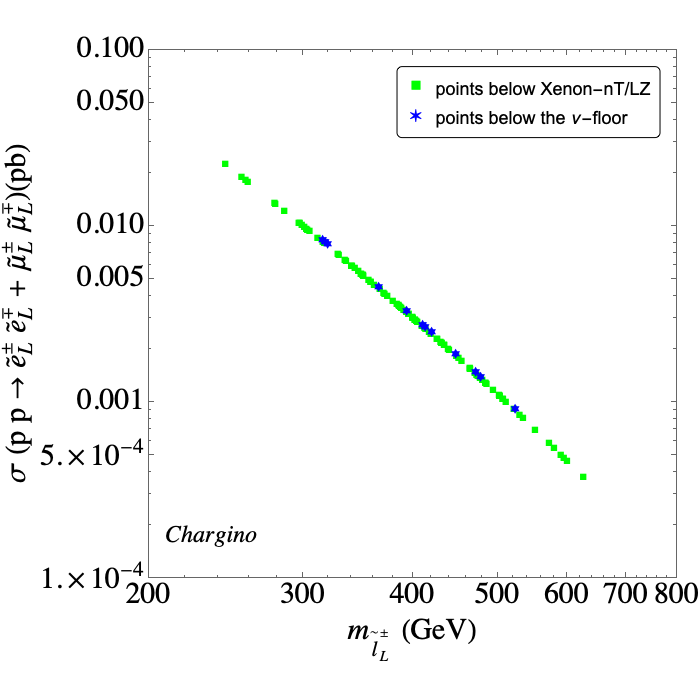} 
\caption{Cross section predictions at  $pp-$ collider with
$\sqrt{s} = 14 \tev$ as a function of the difference of two final state masses.
for the $\cha1$-coannihilation scenario.
The color code indicates the final state, squares are below the
anticipated XENON-nT/LZ reach, stars are below the neutrino floor.
}
\label{fig:ppch-future}
\end{figure}

So far, similar future sensitivity estimates for the slepton pair
production searches by the experimental collaborations are
lacking. However, in order to provide 
an estimate of the production cross section at the HL-LHC,
we compute the NLO+NLL threshold resummed cross sections for
$\tilde e_L^\pm \tilde e_L^\mp$
and $\tilde \mu_L^\pm \tilde \mu_L^\mp$ pair productions for the
bino/wino DM scenario with $\cha1$-coannihilation, 
using the public package
Resummino~\cite{resummino,Bozzi:2006fw,Bozzi:2007qr,Debove:2009ia,Debove:2010kf}.
The result is presented in \reffi{fig:ppch-future}, where in the upper left
plot the production cross section is presented as a function of the
mass difference between the produced particle and the LSP,
$\De m = \mL - \mneu1$,
and in the right plot it is shown as a function of $\mL$. The parameter points
below the reach of XENON-nT/LZ are shown as green squares and those below
the neutrino floor are marked with blue stars.
The production proceeds
through the $s$-channel exchange of $Z$ bosons and photons.
The cross-section for low $\mL$, also roughly corresponding to a
low $\De m$, appear to be significantly large at the level of $\sim 10 \fb$.
However, here it must be taken into account that 
in this case, due to the proximity of $\mcha1$ and $\mneu1$, the sleptons have
a significant $\br(\Slpm \to \nu \chapm1)$, as opposed to the simplified
model assumption of $\br(\Slpm \to l \neu1) = 100$\% . This reduces the
effective cross section to a large extent, making the sleptons harder 
to be probed at the HL-LHC. Consequently, the complementarity
between the DD experiments and the HL-LHC can not conclusively be
answered.

%%%%%%%%%%%%%%%%%%%%%%%%%%%% F I G U R E %%%%%%%%%%%%%%%%%%%%%%%%%%%%%%
%%%%%%%%%%%%%%%%%%%%%%%%%%%%% F I G U R E %%%%%%%%%%%%%%%%%%%%%%%%%%%%%%

%%%%%%%%%%%%%%%%%%%%%%%%%%%%%%%%%%%%%%%%%%%%%%%%%%%%%%%%%%%%%%%%%%%%%%%%%%
\begin{figure}[ht!]
%	\vspace{4em}
\centering
\includegraphics[width=0.45\textwidth]{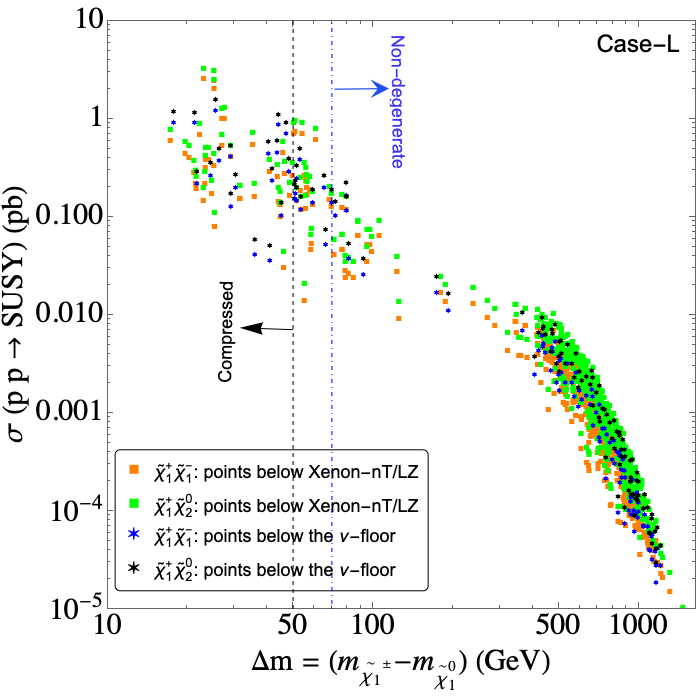}
\includegraphics[width=0.45\textwidth]{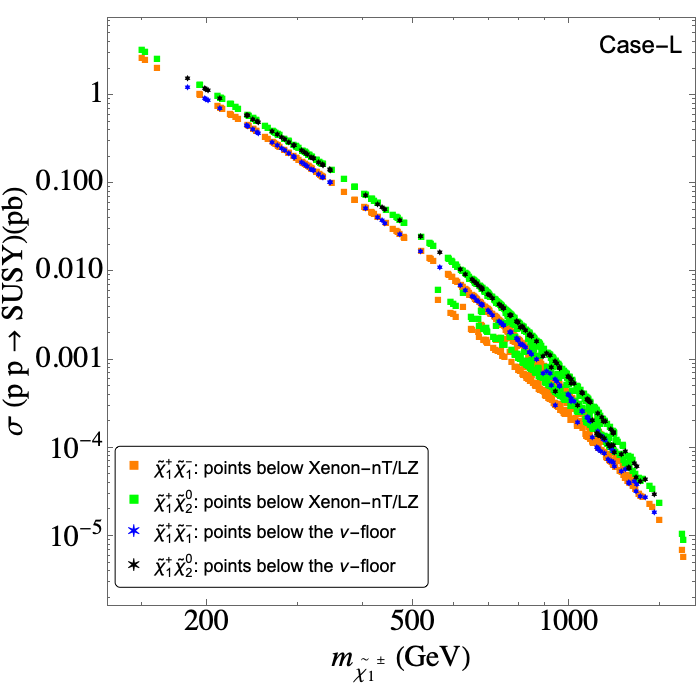} \\
\includegraphics[width=0.45\textwidth]{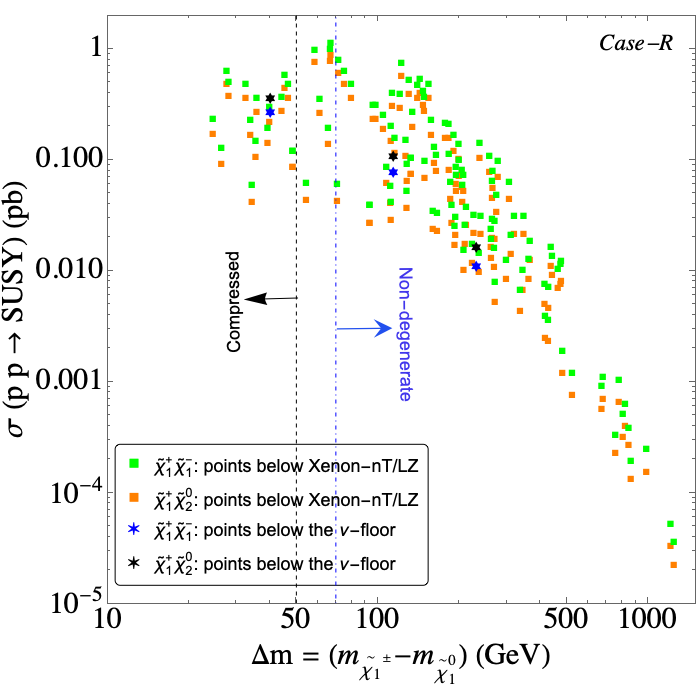}
\includegraphics[width=0.45\textwidth]{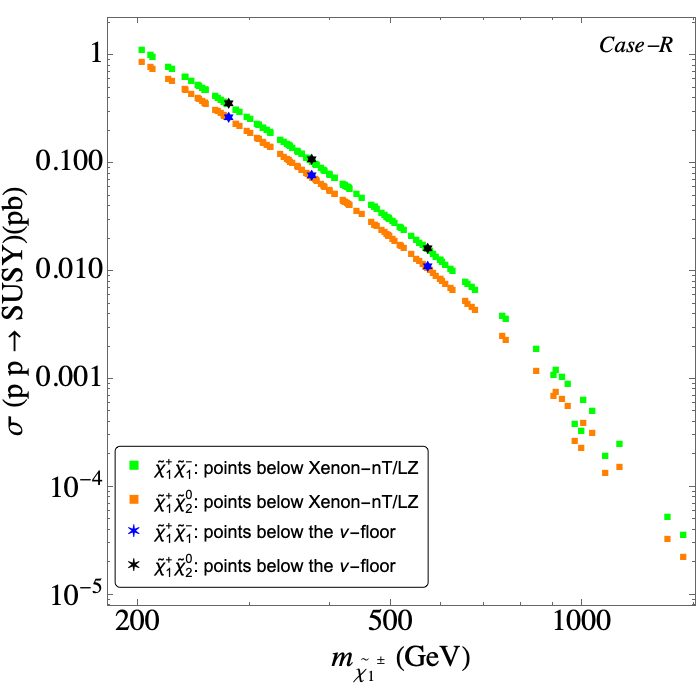} 
\caption{Cross section predictions at  $pp-$ collider with
$\sqrt{s} = 14 \tev$ as a function of $\De m = \mcha1 - \mneu1$
(left) and $\mcha1$ (right).
Upper row: $\Slpm$-coannihilation case-L;
lower row: $\Slpm$-coannihilation case-R.
The color code indicates the final state, squares are below the
anticipated XENON-nT/LZ reach, stars are below the neutrino floor.
}
\label{fig:pp-future}
\end{figure}
%%%%%%%%%%%%%%%%%%%%%%%%%%%% F I G U R E %%%%%%%%%%%%%%%%%%%%%%%%%%%%%%

%%%%%%%%%%%%%%%%%%%%%%%%%%%%%%%%%%%%%%%%%%%%%%%%%%%%%%%%%%%%%%%%%%%%%%%%%%
\begin{figure}[ht!]
%	\vspace{4em}
\centering
\includegraphics[height=65mm,width=0.45\textwidth]{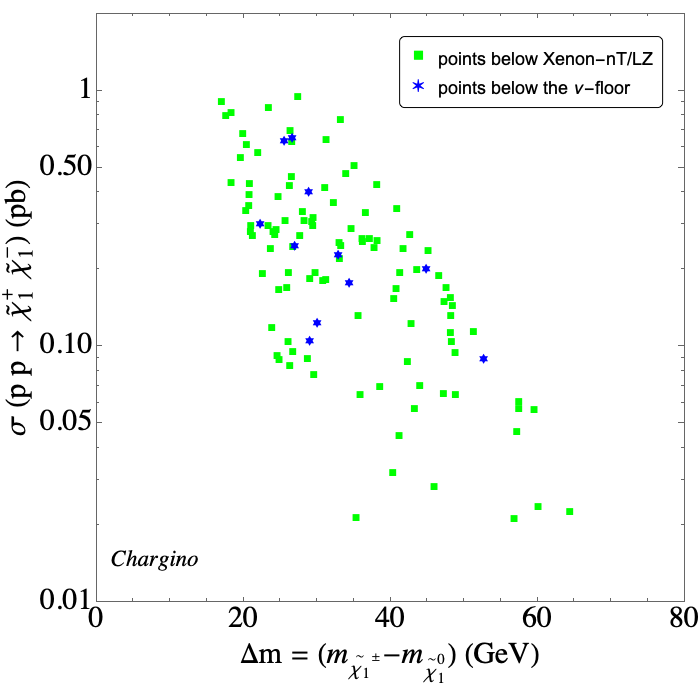}
\includegraphics[height=65mm,width=0.45\textwidth]{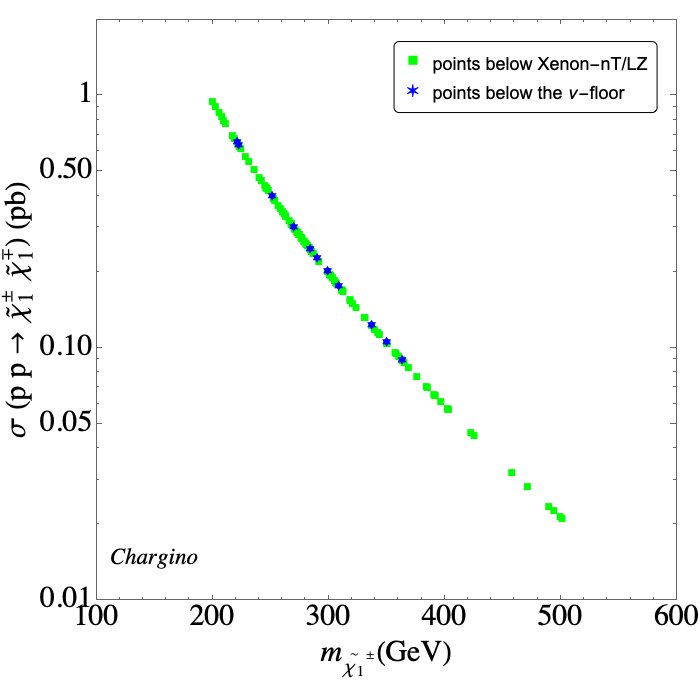} \\
\includegraphics[height=66mm,width=0.45\textwidth]{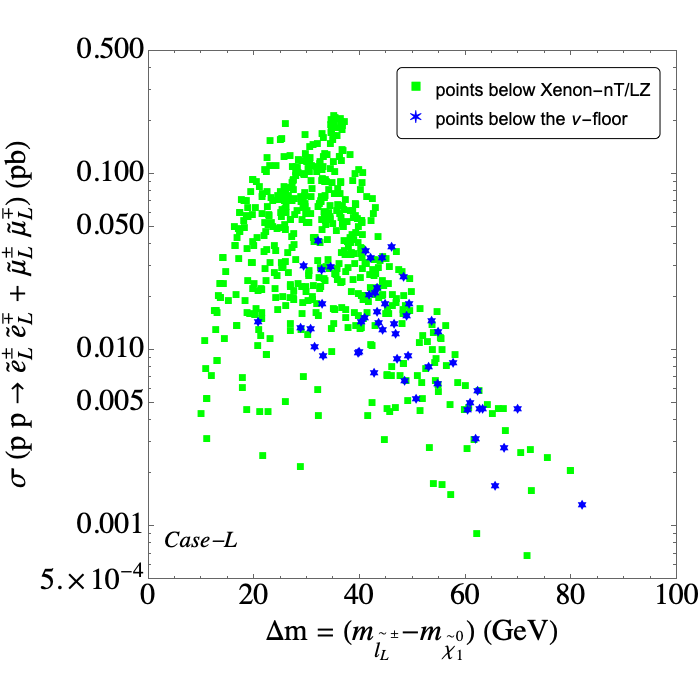}
\includegraphics[height=66mm,width=0.45\textwidth]{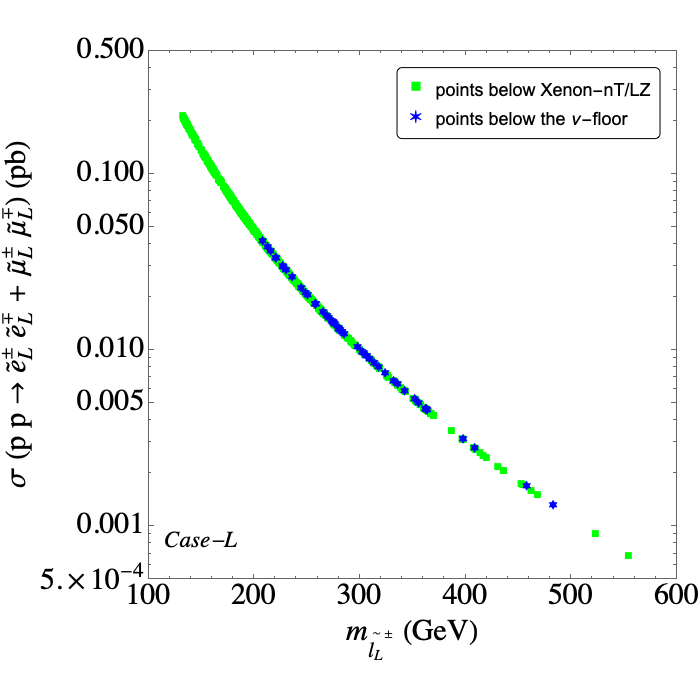} \\
\includegraphics[height=66mm,width=0.45\textwidth]{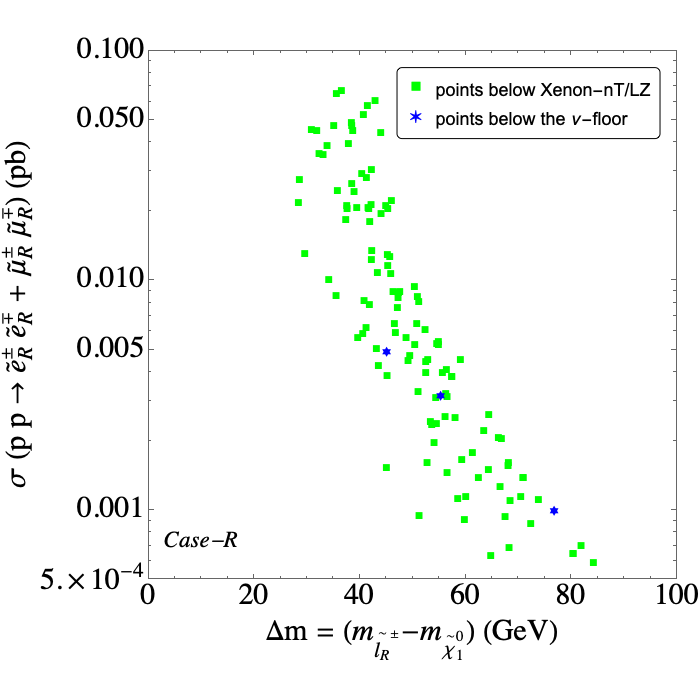}
\includegraphics[height=66mm,width=0.45\textwidth]{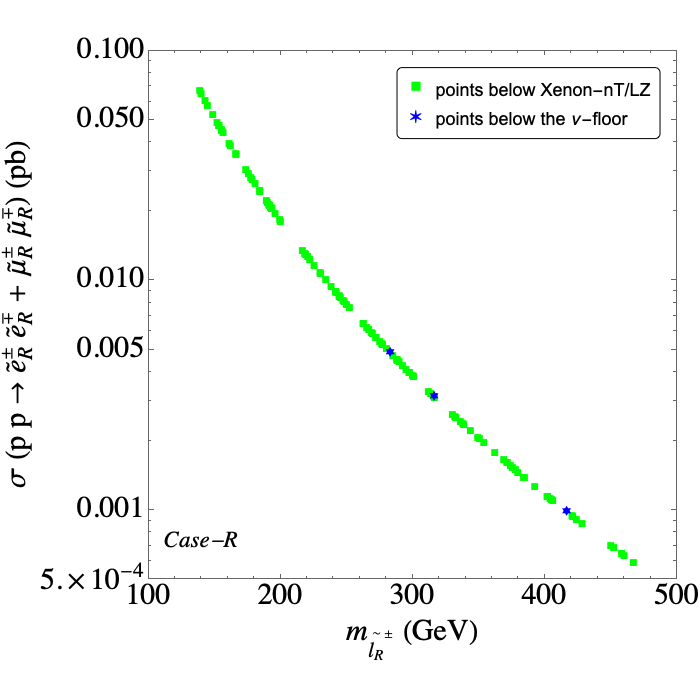} 
\caption{Cross section predictions at  $pp-$ collider with
$\sqrt{s} = 14 \tev$ as a function of  $\De m = m_{\rm NLSP} - \mneu1$ (
left) and $m_{\rm NLSP} $ -  $\mneu1$ (right) for chargino (top),
case-L (middle) and case-R(bottom). 
Green squares are below the
anticipated XENON-nT/LZ reach and blue stars are below the neutrino floor.
}
\label{fig:pp-future-comp}
\end{figure}
%%%%%%%%%%%%%%%%%%%%%%%%%%%% F I G U R E %%%%%%%%%%%%%%%%%%%%%%%%%%%%%%

For the bino DM scenario with $\Slpm$-coannihilation, the searches that
could be the most constraining are those coming from compressed spectra
searches looking for $\Slpm$-pair production, as well as the
$\cha1-\neu2$ production searches leading to three leptons and $\met$ in
the final state. 
No projected sensitivity for the former search exists so far to our knowledge.
For the latter search, the projected 95\% C.L. exclusion contours have been
provided by the ATLAS collaboration~\cite{CidVidal:2018eel}
for the decays 
$\chapm1 \neu2 \to W^\pm Z$  and $\chapm1 \neu2 \to W^\pm h$.
The limits are given for simplified model scenarios assuming $\cha1$ and
$\neu2$ to be 
purely wino-like and mass-degenerate and $\neu1$ to be purely bino-like.
These searches are most effective in the large mass splitting regions,
$\Delta m = \mcha1 - \mneu1 \gtrsim \MZ$ and  $\Delta m \gtrsim \Mh$
for the $W^\pm Z$  and $W^\pm h$ modes, respectively, where they can
probe masses up to $\mcha1 = \mneu2 \sim 1.2 \tev$. 
The parameter region where $\mcha1, \mneu2 > \mL,\mR$, the
$\chapm1, \neu2$ may also decay via sleptons of the first two
generations. The prospect for such decay channels, however has not
been analyzed. In \reffi{fig:pp-future} we show 
our results for the relevant gaugino-pair production cross sections in the
bino DM with $\Slpm$-coannihilation scenarios 
case-L (top row) and case-R (bottom row)
derived at the NLO+NLL accuracy using Resummino.
The squares and stars represent points below the sensitivity of XENON-nT/LZ
and the neutrino floor, respectively. In the left plots we show the
cross sections with respect to the mass difference $\De m = \mcha1 - \mneu1$,
indicating the regions corresponding to compressed and
non-degenerate spectra. In the right plots the cross sections are
shown as a function of $\mcha1$ directly.
As in the case of $\chapm1$-coannihilation, the squarks are assumed to be
very heavy in this case. Thus, the dominant production processes occur via
the $s$-channel exchange of $W, Z$ bosons and photons.
The larger cross section in the  low $\Delta m$ regions may
be beneficial for compressed spectra searches looking for
$\chapm1-\neu2$ pair production. 
In the higher $\Delta m$ region, the cross section decreases steadily
upto $\mcha1 \sim 1.2 \tev$. As in the previous case, also here the apparently
large cross section reached for relatively light $\chapm1$ should
be interpreted with caution in deriving future
exclusion/discovery potentials: on the one hand,
$\chapm1, \neu2$ may decay partly via sleptons of the first two generations,
weakening the limits from gauge-boson or Higgs-mediated decays. On the
other hand, they may decay to some extent via $\stau$'s, relaxing the
bounds from both slepton-mediated and gauge/Higgs-boson mediated decays.
As before, the complementarity
between the DD experiments and the HL-LHC can not conclusively be
answered.

For the sake of completeness, we also show the production
cross-section for the NLSPs in all three cases in
Fig.~\ref{fig:pp-future-comp} as a function of
$\De m(= m_{\rm NLSP} - \mneu1)$ as well as a function of
$m_{\rm NLSP}$. The production cross-section at $14 \tev$
at NLO+NLL is the largest for the chargino co-annihilation reaching
upto \order{1 \pb} for the minimum mass gap between the NLSP and LSP,
while for case-L and case-R, it remains at least one order
below. These pair production of chargino or slepton NLSPs corresponds
to the compressed spectra searches at the HL-LHC where the final state
signal comprises of  ISR jets plus missing energy. However, future
linear colliders will have better sensitivity to probe these signal
regions. The details are discussed in the following section.

%%%%%%%%%%%%%%%%%%%%%%%%%%%%%%%%%%%%%%%%%%%%%%%%%%%%%%%%%%%%%%%%%%%%%%%%%%%%%%

\subsection{ILC/CLIC prospects}
\label{sec:future-ee}

Direct production of EW particles at $e^+e^-$ colliders requires a
sufficiently high center-of-mass energy, $\sqrt{s}$. Consequently,
we focus here on the two proposals for linear $e^+e^-$ colliders,
ILC~\cite{ILC-TDR,LCreport} and CLIC~\cite{CLIC,LCreport}, which can
reach energies up to $1 \tev$, and $3 \tev$, respectively. The former
one we also denote as ILC1000. 
We evaluate the cross-sections for the various LSP and NLSP pair
production modes for $\sqrt{s} = 1 \tev$, which can be reached in the
final stage of the ILC or are below the anticipated CLIC energies
(where at higher $\sqrt{s}$ larger cross sections are obtained).
At the ILC1000 an integrated luminosity of $8 \iab$ is
foreseen~\cite{Barklow:2015tja,Fujii:2017vwa}. 
The cross-section predictions are based on tree-level results, obtained as
in~\citeres{Heinemeyer:2017izw,Heinemeyer:2018szj}. There it was shown
that the full one-loop corrections can amount up to 10-20\%\,%
\footnote{Including the full one-loop corrections here as done
in \citeres{Heinemeyer:2017izw,Heinemeyer:2018szj} would have required to
determine the preferred renormalization scheme for each
point individually (see \citere{Fritzsche:2013fta} for details), which
goes beyond the scope of this analysis.}%
. Here we do not attempt a rigorous experimental analysis,
but follow
analyses~\cite{Berggren:2013vna,PardodeVera:2020zlr,Berggren:2020tle} that
indicate that to a good approximation
final states with the sum of the masses smaller than the
center-of-mass energy can be detected. 

%%%%%%%%%%%%%%%%%%%%%%%%%%%% F I G U R E %%%%%%%%%%%%%%%%%%%%%%%%%%%%%%
\begin{figure}[htb!]
%	\vspace{4em}
\centering
\includegraphics[width=0.45\textwidth]{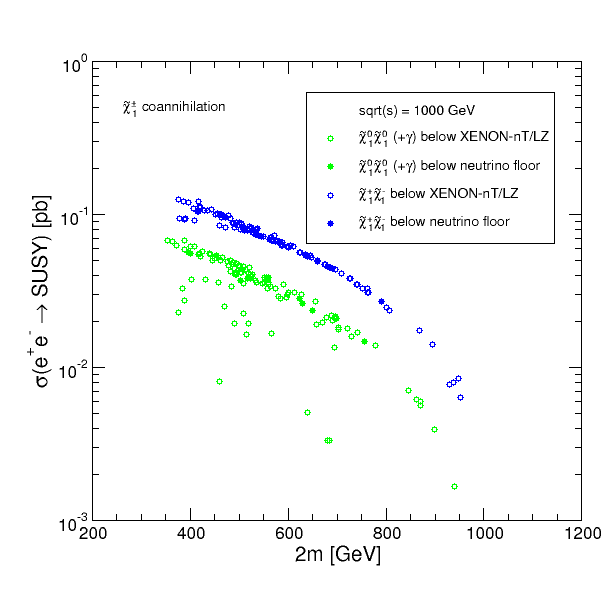}\\
\includegraphics[width=0.45\textwidth]{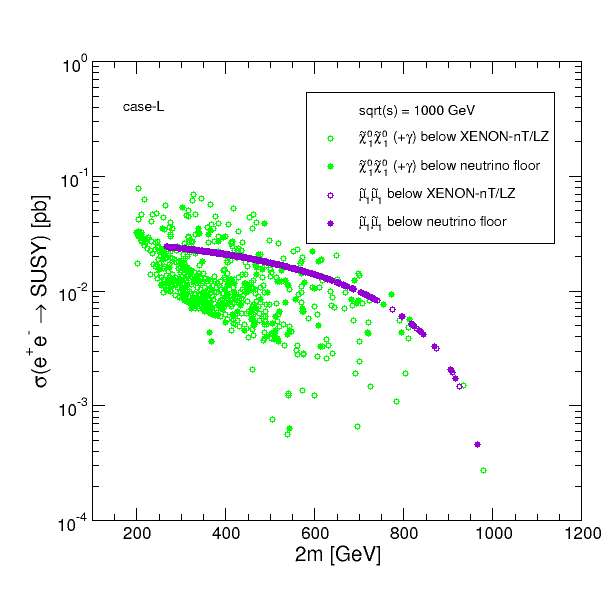}
\includegraphics[width=0.45\textwidth]{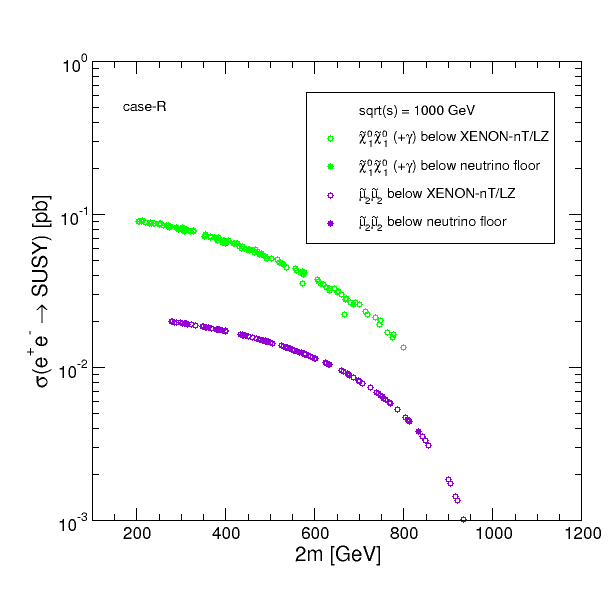}
\caption{Cross section predictions at an $e^+e^-$ collider with
$\sqrt{s} = 1000 \gev$ as a function of the sum of two final state masses.
Upper plot: $\cha1$-coannihilation scenario;
lower left plot: $\Slpm$-coannihilation case-L;
lower right plot: $\Slpm$-coannihilation case-R.
The color code indicates the final state, open circles are below the
anticipated XENON-nT/LZ reach, full circles are below the neutrino floor.
}
\label{fig:ee-future}
\end{figure}
%%%%%%%%%%%%%%%%%%%%%%%%%%%% F I G U R E %%%%%%%%%%%%%%%%%%%%%%%%%%%%%%

In \reffi{fig:ee-future} we show the LSP and NLSP pair production cross
sections for an $e^+e^-$ collider at $\sqrt{s} = 1000 \gev$ as a
function of the two (identical) final state masses. The upper plot shows
$\sig(e^+e^- \to \neu1\neu1 (+ \ga))$ production%
\footnote{
Our tree level calculation does not include the photon radiation, which
appears only starting from the one-loop level. However, such an ISR
photon is crucial to detect this process due to the invisible final
state. We take our tree-level cross section as a rough approximation of
the cross section including the ISR photon, see
also \citere{Heinemeyer:2017izw} and use the notation ``$(+\ga)$''. 
}%
~in green, and
$\sig(e^+e^- \to \chap1\cham1)$ in blue. The open circles are the points
below the anticipated XENON-nT/LZ limit, whereas the solid circles are
the points below the neutrino floor. On can observe that {\it all}
points are within the reach of the ILC1000. The cross sections range
roughly from $\sim 100 \fb$ for low masses to $\sim 10 \fb$ for larger
masses, with only a very few points have smaller cross
sections. Overall, assuming an integrated luminosity of $8 \iab$, this
corresponds to $\sim 80000 - 800000$ events. Consequently, in
contrast to the HL-LHC, the $e^+e^-$
colliders show a clear and conclusive complementarity to the future DD
experiments. The $\cha1$-coannihilation scenario will be fully covered
by either DD experiments or by searches at the ILC1000.

The lower plots of \reffi{fig:ee-future} show the LSP and NLSP
production cross section in the $\Slpm$-coannihilation scenario for
case-L (left) and case-R (right). The green points show again
$\sig(e^+e^- \to \neu1\neu1 (+\ga))$, whereas the violet points left
and right  show 
$\sig(e^+e^- \to \Smu1\Smu1)$ (case-L) and
$\sig(e^+e^- \to \Smu2\Smu2)$ (case-R), respectively. Open and full
circles denote, as above, the points below the anticipated XENON-nT/LZ
limit and the neutrino floor.
The visible spread in the $\neu1\neu1 (+\ga)$ production for case-L
w.r.t.\ case-R is a result of the more complex structure of the
$e^\pm$-$\Sel{L}$-$\neu1$ coupling as compared to the 
$e^\pm$-$\Sel{R}$-$\neu1$ coupling, dominating the $t$-channel exchange
diagram, respectively.
In both cases we see that, as for the $\cha1$-coannihilation case,
all points result in particles that can be pair produced at the
ILC1000 (except the very highest mass points in case-L). The cross
sections range between $100 \fb$ to $10 \fb$ for $\neu1\neu1 (+\ga)$,
and between $20 \fb$ to $1 \fb$ for smuon pair production.
Even for the smallest production cross section this
corresponds to $\sim 8000$ events in the foreseen $1000 \gev$ ILC run.
Also these two cases can conclusively
be probed in the conjunction of DD experiments and an $e^+e^-$ collider
at $\sqrt{s} = 1000 \gev$, in contrast to the HL-LHC, where the
prospects are less clear, see the previous subsection.

%%%%%%%%%%%%%%%%%%%%%%%%%%%%%%%%%%%%%%%%%%%%%%%%%%%%%%%%%%%%%%%%%%%%%%%%%%
%%%%%%%%%%%%%%%%%%%%%%%%%%%%%%%%%%%%%%%%%%%%%%%%%%%%%%%%%%%%%%%%%%%%%%%%%%

\section{Conclusions}
\label{sec:conclusion}

We performed an analysis for the DM predictions of the EW
sector of the MSSM, taking into account all
relevant theoretical and experimental constraints. 
The experimental results comprised the direct searches at the
LHC, the current DM relic abundance (either as an upper limit or as a direct
measurement), the DM direct detection (DD) experiments and in
particular the newly confirmed deviation of the anomalous magnetic moment
of the muon~\cite{Abi:2021gix}. 
As we had analyzed previously~\cite{CHS1,CHS2,CHS3}, five different
scenarios can be classified by the mechanism
that brings the LSP relic density into agreement with the measured
values. These are
{\it (i)} 
higgsino DM ($\mu < M_1, M_2, \msl{L}, \msl{R}$),
DM relic density is only an upper bound (the correct relic density implies
$\mneu1 \sim 1 \tev$ and \gmin2\ cannot be fulfilled), 
$m_{\rm (N)LSP} \lsim 500 \gev$ with $m_{\rm NLSP} - m_{\rm LSP} \sim 5 \gev$;
{\it (ii)}
wino DM ($M_2 < M_1, \mu, \msl{L}, \msl{R}$),
DM relic density is only an upper bound, (the correct relic abundance implies
$\mneu1 \sim 3 \tev$ and \gmin2\ cannot be fulfilled),
$m_{\rm (N)LSP} \lsim 600 \gev$ with $m_{\rm NLSP} - m_{\rm LSP} \sim 0.3 \gev$.
{\it (iii)}
bino/wino DM with $\cha1$-coannihilation ($M_1 \lsim M_2$), correct
DM relic density can be achieved, $m_{\rm (N)LSP} \lsim 650\, (700) \gev$;
{\it (iv)}
bino DM with $\Slpm$-coannihilation case-L ($M_1 \lsim \msl{L}$),
DM relic density can be fulfilled, $m_{\rm (N)LSP} \lsim 650\, (700) \gev$;
{\it (v)}
bino DM with $\Slpm$-coannihilation case-R ($M_1 \lsim \msl{R}$),
DM relic density can be fulfilled, $m_{\rm (N)LSP} \lsim 650\, (700) \gev$;

In this letter we addressed the status of the implications of the new
result for $\De\amu$ (in conjunction with the other constraints) for
the DM predictions in the five scenarios. In a first 
step we analyzed the predictions for the DM relic density as a
function of the (N)LSP masses.
For higgsino and wino DM we analyzed the case where the  $\neu1$
satisfies only a part of the total DM content 
while being consistent with $\De\amu$. On the contrary, for
bino/wino DM and the two bino DM cases, the $\neu1$ LSP can yield
the total DM relic abundance, or only a part of the total DM
content (with the relic abundance limit taken as an upper bound).
As evident, for the heavier mass region of the LSP, significant 
coannihilation is necessary to achieve the relic abundance leading to the
smallest mass gap between the NLSP and LSP for all these
cases. However, for higgsino and wino DM, the NLSP-LSP mass gap is inherently smaller than the other three cases that results in a much compressed spectra.

In a second step we evaluated the prospects for future DD experiments
in the five scenarios. We observed that higgsino and wino DM can be
covered by the ``next round of DM DD experiments'', where we showed
explicitly the anticipated reach of XENON-nT, LZ, DarkSide and
Argo. XENON-nT and LZ have a similar reach, which is moderately improved
by DarkSide and a little more by Argo. For higgsino and wino DM all allowed
points are well in the reach of XENON-nT/LZ.
Therefore, besides the compressed spectra searches at the future hadron
and lepton collider, the future DD experiments are also capable of
testing these scenarios conclusively.
For slepton coannihilation case-L and case-R, the
allowed points with the correct relic abundance are above the
projected reach of XENON-nT/LZ, whereas for  bino/wino DM with
chargino coannihilation a few points also fall within the higher
anticipated sensitivity of DarkSide.  
For lower relic abundances the $\ssi$ values decrease further for these
three scenarios and can go even below the neutrino floor. However, 
in the case of bino case-R DM, the points with the lowest $\ssi$ are
found only marginally below the neutrino floor
and thus can potentially be covered by further future DD experiments.

In continuation, we show that the HL-LHC and the future
$e^+e^-$ collider operating at an energy of up to $1000 \gev$, i.e.\ the
ILC1000 or CLIC can play the complementary role to probe the parameter
space obtained 
below the anticipated XENON-nT/LZ limit or even below the neutrino floor.
For the HL-LHC we focused on the production of the EW particles which
are neither the LSP nor the NLSP, i.e.\ that are not necessarily part of
a compressed EW spectrum. While partially sizable cross sections are
found at the HL-LHC with $\sqrt{s} = 14 \tev$, in particular in the
lower mass ranges, a
proper estimation of the future reach including 
the complex decay structure of the signal region is
mandatory (which often so far are not available) to make
a conclusive judgement. On the other hand,
in the higher mass range, the EW SUSY production cross sections sharply drops
below fb order, specifically for $\Slpm$-coannihilation. Consequently,
it appears unlikely that the points that may escape the DD experiments
can fully be probed at the HL-LHC. For completeness we also calculated
the  production cross-sections
for the compressed spectra searches at the HL-LHC,
where again detailed analyses are not yet available.

The situation is substantially better in the case of an $e^+e^-$
collider with $\sqrt{s} \lsim 1 \tev$. It was shown that at the ILC or
CLIC mass spectra with very small mass splitting can be detected, i.e.\
one does not have to rely on the production of heavier SUSY particles,
but can study the production of the LSP (with an ISR photon) and the
NLSP. We have calculated the corresponding production cross sections for
all points below the XENON-nT/LZ limit or the neutrino floor. It was
shown that effectively the whole parameter space that may escape the DD
experiments can be covered by ILC1000/CLIC.
This demonstrates the important complementarity of
DD experiments and future (linear) $e^+e^-$ colliders to cover the EW
sector of the MSSM.

%%%%%%%%%%%%%%%%%%%%%%%%%%%%%%%%%%%%%%%%%%%%%%%%%%%%%%%%%%%%%%%%%%%%%%%%%%
%\clearpage

\subsection*{Acknowledgments}

We thank
D.~Cerde\~no
and
G.~Moortgat-Pick
for helpful discussions.
I.S.\ thanks S.~Matsumoto for the cluster facility.
The work of I.S.\ is supported by World Premier
International Research Center Initiative (WPI), MEXT, Japan. 
The work of S.H.\ is supported in part by the
MEINCOP Spain under contract PID2019-110058GB-C21 and in part by
the AEI through the grant IFT Centro de Excelencia Severo Ochoa SEV-2016-0597.
The work of M.C.\ is supported by the project AstroCeNT:
Particle Astrophysics Science and Technology Centre,  carried out within
the International Research Agendas programme of
the Foundation for Polish Science financed by the
European Union under the European Regional Development Fund.

%%%%%%%%%%%%%%%%%%%%%%%%%%%%%%%%%%%%%%%%%%%%%%%%%%%%%%%%%%%%%%%%%%%%%%%%%%
%%%%%%%%%%%%%%%%%%%%%%%%%%%%%%%%%%%%%%%%%%%%%%%%%%%%%%%%%%%%%%%%%%%%%%%%%%

%\pagebreak

\newcommand\jnl[1]{\textit{\frenchspacing #1}}
\newcommand\vol[1]{\textbf{#1}}

\newpage{\pagestyle{empty}\cleardoublepage}

%%%%%%%%%%%%%%%%%%%%%%%%%%%%%%%%%%%%%%%%%%%%%%%%%%%%%%%%%%%%%%%%%%%%%%%%%%%

\end{document}